\documentclass[11pt]{article}
\usepackage{amssymb}
\usepackage{amsmath}
\usepackage{amscd}
\usepackage{latexsym}
\usepackage{graphics}
\usepackage{color}
\usepackage{mathtools}
\usepackage{tikz}
\usepackage{hyperref}
\usetikzlibrary{arrows}
\input xy
\xyoption{all}

\catcode`@=11
\renewcommand{\section}
{\@startsection{section}{1}{0pt}{\medskipamount}{\medskipamount}{\large\bf}}
\makeatletter\renewcommand{\subsection}
{\@startsection{subsection}{2}{\z@}{-3.25ex plus -1ex minus -.2ex}
{1.5ex plus .2ex}{\it }}

\numberwithin{equation}{section}
\catcode`@=12

\def\a{\alpha}
\def\b{\beta}
\def\g{\gamma}

\def\e{\epsilon}
\def\l{\lambda}

\def\vp{\varphi}

\def\m{\mu}
\def\n{\nu}

\def\beq{\begin{equation}}
\def\eeq{\end{equation}}
\def\bea{\begin{eqnarray}}
\def\eea{\end{eqnarray}}

\renewcommand{\e}{\,\mathrm{e}\,}
\newcommand{\id}{{\mbf 1}}
\newcommand{\im}{\,\mathrm{i}\,}
\newcommand{\diff}{\mathrm{d}}

\newcommand{\R}{{\mathbb{R}}}

\newcommand{\C}{{\mathbb{C}}}
\newcommand{\Z}{{\mathbb{Z}}}

\newcommand{\Ecal}{{\cal E}}

\newcommand{\Ncal}{{\cal N}}

\newcommand{\Gcal}{{\cal G}}

\newcommand{\yb}{{\bar{y}}}

\newcommand{\ab}{{\bar{a}}}

\newcommand{\ca}{{\cal{A}}}
\newcommand{\cf}{{\cal{F}}}

\newcommand{\man}{{\cal M}}

\newcommand{\mon}{{\cal O}}

\newcommand{\su}{{{\rm SU}(2)}}
\newcommand{\uo}{{{\rm U}(1)}}

\newcommand{\mbf}[1]{{\boldsymbol {#1} }}

\def\Hom{{\rm Hom}}

\def\>{\rangle}
\def\<{\langle}
\def\+{\dagger}
\def\={\ =\ }

\newcommand{\lb}{\left(}
\newcommand{\rb}{\right)}
\newcommand{\beb}{\bar \beta}
\newcommand{\com}[2]{\big[#1,#2\big]}
\newcommand{\hf}{\phi}

\newcommand{\ggc}{\widehat{\mathcal{G}}_{\C}}
\newcommand{\lc}{\Gamma}
\newcommand{\et}{\tilde{e}}
\def\ang1{$15$} 
\def\ang2{$15$}
\def\haken{\mathbin{\hbox to 6pt{%
\vrule height0.4pt width5pt depth0pt
\kern-.4pt
\vrule height6pt width0.4pt depth0pt\hss}}}
\begin{document}

\begin{titlepage}
\setcounter{page}{0}
\begin{flushright}
ITP--UH--02/16\\
EMPG--16--01\\
\end{flushright}

\vskip 1.8cm

\begin{center}

{\Large\bf Sasakian quiver gauge theories and instantons on the conifold}

\vspace{15mm}

{\large Jakob C. Geipel${}^1$}, \ \ {\large Olaf Lechtenfeld${}^1$}, \ \ {\large Alexander D. Popov${}^{1}$}
\ \ and \ \ {\large Richard J. Szabo${}^2$}
\\[5mm]
\noindent ${}^1${\em Institut f\"ur Theoretische Physik}\\ and 
{\em Riemann Center for Geometry and Physics}\\
{\em Leibniz Universit\"at Hannover}\\
{\em Appelstra\ss e 2, 30167 Hannover, Germany}\\
Email: {\tt jakob.geipel@itp.uni-hannover.de , lechtenf@itp.uni-hannover.de , alexander.popov@itp.uni-hannover.de}
\\[5mm]
\noindent ${}^2${\em Department of Mathematics, Heriot-Watt University\\
Colin Maclaurin Building, Riccarton, Edinburgh EH14 4AS, U.K.}\\
and
{\em Maxwell Institute for Mathematical Sciences, Edinburgh, U.K.}\\
and
{\em The Higgs Centre for Theoretical Physics, Edinburgh, U.K.}\\
{Email: {\tt R.J.Szabo@hw.ac.uk}}

\vspace{15mm}

\begin{abstract}
\noindent
We consider $\mathrm{Spin}(4)$-equivariant dimensional reduction of Yang-Mills theory on manifolds of the 
form $M^d \times T^{1,1}$, where $M^d$ is a smooth manifold and
$T^{1,1}$ is a five-dimensional Sasaki-Einstein manifold $\mathrm{Spin}(4)/\uo$. We obtain new quiver gauge theories on $M^d$ 
extending those induced via reduction over the leaf spaces $\C P^1 \times \C P^1$ in $T^{1,1}$. We describe 
the Higgs branches of these quiver gauge theories as moduli spaces of $\mathrm{Spin}(4)$-equivariant instantons on the 
conifold which is realized as the metric cone over $T^{1,1}$. We
give an
explicit construction of these moduli spaces as K\"ahler quotients. 
\end{abstract}

\end{center}
\end{titlepage}

{\baselineskip=11pt
\tableofcontents
}

\bigskip

\section{Introduction} 

The idea of extra dimensions has become an important concept in
physics, particularly in string theory wherein the compactification of these dimensions is a 
fundamental ingredient. In this approach one studies theories living on the product $M^d \times \mathbb{X}$ of a $d$-dimensional 
spacetime $M^d$ and a Riemannian manifold $\mathbb{X}$. The latter
space parameterizes the internal degrees of freedom and is usually chosen with reduced
holonomy. While Calabi-Yau manifolds are particular examples, one
faces an enormous number of possible geometric structures with each of them leading to a different 
effective theory on spacetime upon dimensional reduction.

Because of their intensive treatment in differential geometry and their
symmetries, coset spaces $\mathbb{X} = G/H$ are typical candidates for the description of 
the internal degrees of freedom; dimensional reduction over these spaces
is known as \emph{coset space dimensional reduction} \cite{Coset92}. If one considers Yang-Mills theory 
on these spaces and imposes a \mbox{$G$-equivariance} condition on the pertinent bundles and connections, 
systematic restrictions follow and the effective field theories can be
described as quiver gauge theories. The field content constitutes
representations of certain quivers, which are oriented graphs whose arrow representatives can be interpreted as Higgs fields. 
A rigorous mathematical treatment can be found in \cite{GP02}, while
brief reviews can be found in e.g.~\cite{LPS07, DS11}. 

Typical coset spaces $\mathbb{X}$ that have been studied in the
literature are homogeneous spaces carrying K\"ahler structures such as
the complex projective line
$\C P^1$ \cite{SL2C,PS06,Bismas,Dolan:2009ie,Szabo:2014zua} and $\C P^1 \times \C P^1$ \cite{LPS06}, or K\"ahler manifolds of the form $\mathrm{SU}(3)/H$ \cite{SU3,Dolan:2009nz}. Since Sasakian geometry is the natural odd-dimensional counterpart of
K\"ahler geometry, one may include five-dimensional Sasakian
manifolds in this framework of quiver gauge theory \cite{superconformal05}. 
In particular Sasaki-Einstein manifolds $\mathbb{X}$, whose metric
cones $C(\mathbb{X})$ are
Calabi-Yau threefolds, find applications in string theory where they provide
explicit tests of AdS/CFT duality. In this setting the
near horizon geometry of a stack of D3-branes is that of
$AdS_5\times\mathbb{X}$, and the supergravity D3-brane solution interpolates
between $AdS_5\times\mathbb{X}$ and $\R^{1,3}\times C(\mathbb{X})$. In the
low-energy limit, the worldvolume theory on the D-branes thus gives rise to a
superconformal quiver gauge theory in four dimensions which is the (naive)
dimensional reduction of ten-dimensional $\mathcal{N}=1$
supersymmetric Yang-Mills theory over the cone $C(\mathbb{X})$.

As pointed out in \cite{Sparks10}, any complete homogeneous
Sasaki-Einstein manifold of dimension five is a $\uo$-bundle over
either the complex projective plane
 $\C P^2$ or $\C P^1 \times \C P^1$, which respectively realize the two most prominent examples: the five-sphere $S^5$ and the space $T^{1,1}$. In general, 
the notation $T^{p,q}$ \cite{Romans85, Gubser} refers to a class of homogeneous spaces $\mathrm{Spin}(4)/\uo = \su \times \su/\uo$, where 
the coprime integers $p$ and $q$ parameterize the embedding of $\uo$ and, equivalently, 
the Chern numbers $(p,q)\in H^2(\C P^1 \times \C P^1,\Z)$ of the circle bundle. The case $p=1=q$ is best known for the fact that its metric cone is the \emph{conifold}, 
which has been intensively studied both in mathematics and
string theory. Much attention has been paid to the dual
$\mathcal{N}=1$ superconformal quiver gauge theories \cite{T11,
  domainwall} and to configurations of branes probing its conical singularity~\cite{D-Branes,Susy_conifold}, as well as 
to deformations and (partial) resolutions thereof~\cite{GeomTrans, CO1990}. New classes of 
Sasaki-Einstein structures on $S^2\times S^3$, denoted $Y^{p,q}$, have
been constructed in \cite{SaEin_S2S3} which contain the homogeneous
space $T^{1,1}=Y^{1,0}$ as a special case~\cite{Martelli:2004wu}; these spaces have been 
studied in \cite{superconformal05} in the context of their dual
superconformal quiver gauge theories.

Manifolds with special geometry, such as those with reduced holonomy or
$G$-structures, are of interest as backgrounds in string theory due to the 
benefits the additional geometric structures provide for the
construction of explicit solutions. In this context it is shown
in~\cite{noelle12} that the existence of real Killing spinors,
as is the case for Sasaki-Einstein manifolds, implies that a
generalized instanton condition automatically leads to the Yang-Mills equations of gauge theory. 
Moreover, the generalized definition of an instanton from \cite{noelle12} includes the 
 gaugino Killing spinor equation as one part of the BPS equations in heterotic string theory \cite{heterotic}. 

This article addresses the construction of new quiver gauge theories associated to the space $T^{1,1}$. We will obtain them by imposing $\su \times \su$-equivariance on the connections on
vector bundles over this manifold, in the spirit of~\cite{GP02}. To implement the equivariance condition on the connection explicitly, we use an
established framework \cite{IP12} that is also applied for the investigation of instanton solutions e.g. in \cite{conical14, Bunk14, Lubbe}.
A similar study of the quiver gauge theories for the five-sphere $S^5$
and a class of its lens spaces has been performed in \cite{S5},
extending the treatment of~\cite{Lechtenfeld:2014fza} which dealt
uniformly with all
Sasaki-Einstein three-manifolds; there these field theories were
dubbed \emph{Sasakian quiver gauge theories}. 

This paper is organized as follows. In Section~2 we review the geometric properties of the space $T^{1,1}$ and provide the necessary 
basic tools for our ensuing calculations, in particular the choice of local coordinates and the structure equations. By imposing the 
Sasaki-Einstein condition, all pertinent parameters are fixed. The
canonical connection, which is the starting point for the construction of 
instantons, is also introduced. 
Section~3 reviews the general construction of equivariant connections
and determines the resulting quiver gauge theories for equivariant dimensional
reduction over the coset space $T^{1,1}$. Since this
space is a principal $\uo$-bundle over $\C P^1 \times \C P^1$, our descriptions follow closely those from~\cite{LPS06}. Besides the general form of the quiver gauge theories, we
consider some explicit examples and compare them with the quiver gauge theories obtained from dimensional reduction over the coset
space $\mathbb{X}=\C P^1 \times \C P^1$ from~\cite{LPS06}. We shall find that not only vertex
loop modifications occur in the underlying quivers, as in \cite{Lechtenfeld:2014fza,S5}, but also more general
additional arrows, because the group $H=\uo$ is smaller than the
maximal torus of $G=\su\times\su$ and consequently provides fewer
restrictions. We further compute
the curvatures of equivariant connections, and carry out the dimensional reduction of Yang-Mills theory to~$M^d$. In order to understand the structure of the quiver gauge theory more clearly, 
we consider a special case in which the computations are significantly
simplified due to a grading of the equivariant connections.
In Section~4 we study quiver gauge theory on the metric cone $C(T^{1,1})$ over $T^{1,1}$ and impose the Hermitian Yang-Mills equations
in order to obtain solutions of the generalized instanton equations. The moduli spaces of solutions to the resulting equations for spherically symmetric configurations in this
framework have been analysed in~\cite{MS15,S5} in terms of K\"ahler
quotients and adjoint orbits, and we adapt this analysis to our
setting. We also comment on the relation of this description of the Higgs branches of our quiver
gauge theories to moduli spaces of BPS states of
D-branes wrapping $C(T^{1,1})$. Finally, in Section~5 we close with some concluding
remarks, while an appendix at the end of the paper contains some
technical details involving connections and curvatures which are employed throughout the main text.


\section{Geometry of the coset space $T^{1,1}$}
\subsection{Local geometry}

In this section we shall review the geometry of the coset
space $T^{1,1}$; this geometry is well-known both in the physics
literature \cite{Romans85} and in mathematics literature on Sasakian
geometry, see e.g.~\cite{Sparks10}. A description of the geometry 
of the five-dimensional Stiefel manifold
$V_{4,2}= \mathrm{SO}(4)/\mathrm{SO}(2)= \mathrm{SO}(3)\times \mathrm{SO}(3)/\mathrm{SO}(2)$, which has the same structure as $T^{1,1}$ at the Lie algebra level, can be found e.g. in the classification~\cite{NatRed14}.

We start by describing explicit local coordinates on $\su\simeq S^3 $ and $\C P^1 \simeq S^2$, based on the defining 
representation of the Lie group $\su$ on $\C^2$ and the Maurer-Cartan
form. Each element of $\su$ can  be locally written as\footnote{
This description is based on a treatment of the Hopf fibration $S^3 \rightarrow S^2$ and can 
be found e.g. in \cite{Popov09}.}
\beq
\label{eq1}
 \underbrace{\frac{1}{\lb 1+y_l \, \yb_l\rb^{1/2}} \,
  \begin{pmatrix}
  1	&	-\yb_l	\\
  y_l	&	1
  \end{pmatrix}}_{\= g_l\ \in \ \C P^1\ \subset\ \su} \ 
  \underbrace{
  \begin{pmatrix}
  \e^{\im \vp_l}	&	0	\\
  0			& \e^{-\im \vp_l}
  \end{pmatrix}}_{\in \ \mathrm{U}\lb1\rb} \ ,
\eeq
where $y_l$ and $\yb_l$ are stereographic coordinates on $S^2$, defined as in \cite{LPS06}, and  the index  $l=1,2$ refers
 to the two copies of $S^2$ which are contained in $T^{1,1}$. The canonical flat connection $A_l$ on the homogeneous space 
$\C P^1$ is given by the Maurer-Cartan form
\bea
A_l\ \coloneqq \ g_l^{-1} \, \diff g_l 
   \= \frac{1}{1+y_l \, \yb_l} \, \begin{pmatrix}
                              \frac{1}{2} \lb \yb_l \, \diff y_l
                              -y_l \, \diff \yb_l\rb  & -\diff \yb_l\\
			      \diff y_l
                              & \frac{1}{2} \lb y_l \, \diff \yb_l
                              - \yb_l \, \diff y_l\rb 	
                              \end{pmatrix}
                              \ =: \ \begin{pmatrix}
                                 a_l	&	-\beb_l\\
				 \b_l	&	-a_l
                                 \end{pmatrix}
\label{eq2}
\eea
which provides $\mathrm{SU}(2)$-invariant \mbox{1-forms}
\bea
\label{eq3}
a_l \= - \ab_l \=\mbox{$\frac{1}{2}$} \lb \yb_l \, \b_l - y_l \,
\beb_l\rb \ , \ \ \ \ \ \ \ \b_l \= \frac{\diff y_l}{1+y_l\, \yb_l}
\eea
with differentials
\bea
\label{eq4}
\diff a_l \=  - \b_l \wedge \beb_l \ , \ \ \ \ \
\diff \b_l \=  2 a_l \wedge \b_l \ , \ \ \ \ \ 
\diff \beb_l \=  - 2 a_l \wedge \beb_l \ .    
\eea
Since the geometry of $T^{1,1}$ involves the Hopf fibration, it has
a close relation to quantities associated with magnetic monopoles as 
the appearance of the monopole forms (\ref{eq3}) indicates. 

To deal with two copies of $\su$, we can analogously to (\ref{eq1}) start again from the 
defining representation and express an
arbitrary element of $\su \times \su$ locally as
\bea
\label{eq5}
\underbrace{
\begin{pmatrix}
g_1		& \mbf{0}_{2}\\
\mbf{0}_{2} 	& g_2
\end{pmatrix}}_{\in\ \C P^1 \times \C P^1} \ 
\underbrace{
\mathrm{diag}\lb \e^{\im \vp_1}, \e^{-\im \vp_1}, \e^{\im \vp_2},
\e^{-\im \vp_2} \rb}_{ \ \in \ \uo \times \uo}.
\label{rep:su2}
\eea
To pass to the coset space $T^{p,q}$, we have to factor by the
$\uo$ subgroup whose embedding is described by the coprime integers $p$ and $q$, which sends $z\in\uo$ to ${\rm diag}(z^p,z^{-p},z^{-q},z^q)\in \su\times\su$.
We will specialize to the case $p=q=1$, which means that the embedding of $H=\uo$ into 
$G=\su \times \su$ is such that $H$ is generated by the difference of the two Cartan generators of $G$, i.e. 
$\mathfrak{h}= \langle I_{(1)}^3-I_{(2)}^3\rangle$.\footnote{
Two different descriptions of $T^{1,1}$ occur in the literature: Some
treatments (e.g. \cite{Romans85,T11}) obtain the manifold
$T^{1,1}$ from 
$S^3 \times S^3$ by quotienting with the \emph{sum} of the diagonal
$\su$ generators, whereas others (e.g.~\cite{altT11}) quotient by the
$\uo$ subgroup generated by the 
\emph{difference}. Changing from one convention to the other simply inverts the complex structure on one of the two-spheres $S^2$ contained 
in $T^{1,1}$.} 
Therefore we change $\uo$ coordinates to $\vp = \frac{1}{2} \lb
\vp_1 +\vp_2\rb$ and $\psi = \frac{1}{2} \lb \vp_1-\vp_2\rb$, 
so that the $\uo \times \uo$ factor  in (\ref{rep:su2}) reads
\bea
\label{eq6}
\nonumber &&\mathrm{diag}\big( \e^{\im \lb \vp+\psi\rb},\e^{-\im \lb \vp +\psi\rb},\e^{\im \lb -\vp+\psi \rb},\e^{\im \lb \vp -\psi\rb} \big) \\ 
          &&\ \ \ \ \ \ \ \ \ \ \ \ \ \ \ \ \ \ \ \ \ \ \ \= \mathrm{diag}\lb \e^{\im \vp},\e^{-\im \vp},\e^{\im \vp},\e^{-\im \vp} \rb 
\ \mathrm{diag}\big( \e^{\im \psi},\e^{-\im \psi},\e^{-\im
  \psi},\e^{\im \psi}\big) \  .
\eea
By passing to the coset space $T^{1,1}$, the second term in (\ref{eq6}) is divided out and one ends up with elements of the form
\bea
\label{eq7}
v \=
\begin{pmatrix}
g_1		& \mbf{0}_{2}\\
\mbf{0}_{2} 	& g_2
\end{pmatrix}
\ \mathrm{diag}\lb \e^{\im \vp}, \e^{-\im \vp}, \e^{\im \vp}, \e^{-\im
  \vp}\rb \ .
\eea

Hence the local description of $T^{1,1}$ is based on the
quintuple of CR coordinates $(y_1, \yb_1, y_2, \yb_2, \vp)$, 
and we derive a basis  of  $\mathrm{SU}(2)\times \mathrm{SU}(2)$ left-invariant \mbox{1-forms} by considering its canonical 
flat connection
\bea
A_0 \ \coloneqq \ v^{-1} \, \diff v \=
                    \begin{pmatrix} \im \diff \vp+a_1	 & 	-
                      \e^{-2\im \vp} \, \beb_1 	&	0	&	0\\
                                    \e^{2\im \vp}\, \b_1 		 & 	-\lb \im \diff \vp+a_1 \rb 		&	0	&	0 \\
				    0				 &
                                    0 				&
                                    \im \diff \vp +a_2 & -\e^{-2\im
                                      \vp}\, \beb_2\\
                                    0 				 &
                                    0				&
                                    \e^{2\im\vp}\, \b_2  & -\lb \im \diff \vp +a_2\rb
                    \end{pmatrix} \ .
\label{eq8}
\eea 
By introducing the definitions
\bea
\nonumber a \ \coloneqq\  \mbox{$\frac{1}{2}$} \lb a_1-a_2\rb \ , \ \ \ \
 \im \kappa \, e^5 \ \coloneqq\  \im \diff \vp +\mbox{$\frac{1}{2}$}
 \lb a_1+a_2\rb \ , \\[4pt]
 \alpha_1 \lb e^1 +\im e^2\rb\ \coloneqq\  \e^{2 \im \vp}\, \b_1 \ , \ \ \ \
 \alpha_2 \lb e^3 +\im e^4\rb \ \coloneqq\  \e^{2\im \vp}\, \b_2 \ ,
\label{eq9}
\eea
where $\alpha_1$, $\alpha_2$ and $\kappa$ are real constants to be determined later from the Sasaki-Einstein condition, we obtain the expression
\bea
\label{eq10}
 A_0 \= \begin{pmatrix}
                        \im  \kappa \, e^5 +a	&	 -\alpha_1 \lb e^1-\im e^2\rb 				&	0	&	0\\
			  \alpha_1 \lb e^1+\im e^2\rb
                          &	 -\im  \kappa \, e^5 -a & 0      &0\\
			0
                        &	0				& \im
                        \kappa \, e^5 -a & -\alpha_2 \lb e^3- \im e^4\rb \\
			0
                        &	0				&
                        \alpha_2 \lb e^3+\im e^4\rb  &  -\im \kappa \,
                        e^5 +a
                       \end{pmatrix} \ .
\eea

\subsection{Sasaki-Einstein geometry}
Based on the choice of the basis \mbox{1-forms} $e^1, \ldots, e^5$ on $T^{1,1}$ according to (\ref{eq9}), the structure equations can be determined. 
The equation for $e^5$ follows directly from its definition and the differentials (\ref{eq4}), while the other equations can be obtained from
the flatness condition on the canonical connection, \mbox{$\diff A_0 = -A_0 \wedge A_0$}. Since the forms $a_l$ and consequently also $a$ are purely imaginary, 
one ends up with the equations
\bea
\nonumber \diff e^1 &=& 2 \kappa \, e^{25} -2 \im e^2 \wedge a \ , \ \ \ \ \ \
\diff e^2  \= -2  \kappa \, e^{15} +2\im  e^1 \wedge a \ ,\\[4pt]
\nonumber \diff e^3 &=& 2 \kappa\, e^{45}  + 2 \im e^4 \wedge a \ , \ \ \ \ \ \
\diff e^4  \= -2  \kappa \, e^{35} -2\im e^3 \wedge a \ ,\\[4pt]
\diff e^5 &=& \mbox{$\frac{1}{\kappa}$} \lb \alpha_1^2\, e^{12}
+\alpha_2^2\, e^{34}\rb \ ,
\label{eq13}
\eea
where in general we write $e^{\mu_1\cdots
  \mu_k}:=e^{\mu_1}\wedge\cdots\wedge e^{\mu_k}$.

The remaining scaling factors in the definition of the \mbox{1-forms} in (\ref{eq9}) can be fixed by imposing the Sasaki-Einstein condition. 
Among the numerous definitions concerning contact geometry, we use the description given in \cite{CS}. Then a Sasaki-Einstein five-manifold is
characterized by a special \mbox{$\su$-structure} which can be defined by an orthonormal cobasis $\{e^{\m} \}$ and forms 
\beq
\label{eq14}
\eta \= -e^5 \ , \ \ \ \ \ \omega^1 \= e^{23}+e^{14} \ , \ \ \ \ \
\omega^2 \= e^{31}+e^{24} \ , \ \ \ \ \ \omega^3 \= e^{12}+e^{34} 
\eeq
satisfying the equations
\beq
\label{eq15}
\diff \eta \= 2 \omega^3 \ , \ \ \ \ \ \ \diff \omega^1 \= - 3
\eta\wedge \omega^2 \ , \ \ \ \ \ \ \diff \omega^2 \= 3 \eta \wedge
\omega^1 \ . 
\eeq
Here $\eta$ is the contact 1-form which is a connection on the
Sasakian fibration,\footnote{
The 1-form $\eta$ is dual to the R-symmetry
generator of the AdS/CFT dual superconformal gauge theory.
} and $\omega^3$ is the K\"ahler 2-form
of its base. Calculating the differentials with the help of the structure equations (\ref{eq13}) yields
\bea
\label{eq16}
\nonumber \diff \omega^1 &=&  4 \kappa \lb e^{135}-e^{245}\rb \=  4
\kappa \, \eta \wedge \omega^2 \ ,\\[4pt]
\nonumber \diff \omega^2 &=& 4 \kappa \lb e^{145}+e^{235}\rb \= -4
\kappa \, \eta \wedge \omega^1 \ ,\\[4pt]
             \diff \eta &=& -\diff e^5 \= -\mbox{$\frac{1}{\kappa}$}
             \lb \alpha_1^2\, e^{12}+\alpha_2^2\, e^{34}\rb \ .
\eea
Hence in order to fulfill the Sasaki-Einstein condition one has to impose 
\beq
\label{eq17}
\kappa \= -\mbox{$\frac{3}{4}$} \ , \ \ \ \ \ \ \ \ \ \ \ \ \
\alpha_1^2 \= \alpha_2^2 \= \mbox{$\frac{3}{2}$} \ .
\eeq
With this choice of parameters the definition of the basic \mbox{1-forms} is given by
\bea
\label{eq19}
e^1 +\im e^2 \= \sqrt{\mbox{$\frac{2}{3}$}} \, \e^{2 \im \vp} \, \b_1
\ , \ \ \ \ 
e^3 +\im e^4 \= \sqrt{\mbox{$\frac{2}{3}$}} \, \e^{2 \im \vp} \, \b_2
\ , \ \ \ \
e^5 \= -\mbox{$\frac{4}{3}$}\, \diff \vp + \mbox{$\frac{2\im}{3}$} 
\lb a_1 +a_2\rb \ .
\eea

The implications of the conditions (\ref{eq17}) for the Riemannian
geometry of $T^{1,1}$ is as follows. Since our geometry consists 
of two copies of $\C P^1 \simeq S^2$, the \mbox{round K\"ahler metric \cite{LPS06}}
\beq
g_{S^2 \times S^2} \= 4 R_1^2\;\b_1 \otimes \beb_1 + 4 R_2^2\;\b_2 \otimes \beb_2 
\eeq
parameterized by two radii $R_l$ appears.
In our case, the Sasaki-Einstein condition fixes these radii. 
Recalling the orthonormality of the forms $e^{\m}$, we obtain the metric
\bea
\label{eq20}
 g \= \delta_{\m\n} \, e^{\m} \otimes e^{\n} 
                           \= \mbox{$\frac{2}{3}$}\, \b_1 \otimes
                           \beb_1 +\mbox{$\frac{2}{3} $}\, \b_2
                           \otimes \beb_2 + \eta \otimes \eta
\eea
with implicit summation throughout over repeated upper and lower indices. This shows that imposing this condition requires $R_1^2=R_2^2=\frac{1}{6}$ (see also the metric in \cite{CO1990}). 
In particular, we cannot rescale the radii as for K\"ahler structures on the 
\mbox{coset space $\C P^1 \times \C P^1$}.

\subsection{Canonical connection}
We proceed to the definition of the canonical connection on $T^{1,1}$
and its curvature. 
Recall that the structure equations relate the connection
\mbox{1-forms} $\Gamma^{\m}_{\ \n}$, the torsion 2-form $T^{\m}$, and the
differentials of the basis \mbox{1-forms} $e^\m$ by
\beq
\diff e^{\m} \= -\Gamma^{\m}_{\ \n} \wedge e^{\n}+T^{\m} \ .
\eeq 
Hence the Levi-Civita connection, determined by the requirement
$T^{\m} = 0$, is expressed by the non-vanishing components (see
Appendix~A for details)
\bea
\nonumber \lc^{1}_{\ 2}&=& -\mbox{$\frac{1}{2}$}\, e^5-2 \im a \ , \
\ \ \lc^{1}_{\ 5} \= e^2 \ , \ \ \ \lc^{2}_{\ 5}\= -e^1 \ ,\\[4pt]
          \lc^{3}_{\ 4}&=& -\mbox{$\frac{1}{2}$}\, e^5+2 \im a \ , \
          \ \ \lc^{3}_{\ 5}\= e^4 \ , \ \ \ \lc^{4}_{\ 5}\= -e^3 \ ,
\label{conn_lc}
\eea
with the antisymmetry $\Gamma^{\m}_{\ \n}=-\Gamma^{\n}_{\ \m}$. The curvature of the Levi-Civita connection 
yields the Ricci tensor 
\beq
\mathrm{Ric}_g \= 4 \delta_{\m\n} \, e^{\m} \otimes e^{\n} 
\= 4 g \ ,
\label{ricci_t11}
\eeq
which confirms that the space with the chosen metric is also Einstein. Moreover, the structure has generic holonomy, 
i.e.~the entire Lie algebra $\mathfrak{so}(5)$.

In dealing with special geometries, it is useful to consider adapted connections that are compatible with the given structure. 
Declaring all terms in (\ref{conn_lc}), apart from those containing the form $a$, to be  torsion, we obtain the $\mathrm{U}(1)$ connection
\beq
\Gamma^{1}_{\ 2}\=-2 \im a \ , \ \ \ \Gamma^{2}_{\ 1}\= 2 \im a \ , \
\ \ \Gamma^3_{\ 4}\= 2 \im a \ , \ \ \ \Gamma^4_{\ 3}\=-2 \im a \ ,
\label{u1-connection}
\eeq
which is the \emph{canonical connection} on $T^{1,1}$ viewed as a homogeneous space ${\rm Spin}(4)/\uo$. This connection coincides with the canonical connection on $T^{1,1}$ viewed as a Sasaki-Einstein manifold, as introduced in~\cite{noelle12}.\footnote{
For a Sasaki-Einstein five-manifold, the torsion of the canonical connection is given by~\cite{noelle12}
$T^5=P_{5\m\n}\, e^{\m\n}$ and $T^a=\frac{3}{4}\, P_{a\m\n}\,
e^{\m\n}$ for $a=1,2,3,4$ with $P=\eta \wedge \omega^3$.}  
We shall use the canonical connection as a starting point for our investigation because it is an instanton, 
according to the generalized definition in \cite{noelle12}. For a five-dimensional Sasaki-Einstein manifold,
the instanton equation is given by
\beq
\star R \= -\star Q \wedge R \qquad\textrm{with}\qquad
Q\=\mbox{$\frac{1}{2}$}\, \omega^3 \wedge \omega^3\=e^{1234}
\label{insta:eq}
\eeq
for a curvature 2-form $R$, where $\star$ is the Hodge operator associated to the Sasaki-Einstein metric.
The curvature of the $\mathrm{U}(1)$ connection (\ref{u1-connection}) reads
\beq
R^{1}_{\ 2}\= -R^{2}_{\ 1}\= 3 \lb e^{12} -e^{34} \rb \= R^{4}_{\ 3}\= -R^{3}_{\ 4}
\label{curv_lc}
\eeq
and (after rescaling) it indeed solves the instanton equation $(\ref{insta:eq})$. 

\section{Quiver gauge theory}

\subsection{Quiver bundles}
Since the Sasaki-Einstein manifold in our discussion is realized as a coset space $G/H$, 
a very natural condition to impose is $G$-equivariance
of the vector bundles over $T^{1,1}$ which carry a gauge connection. A detailed mathematical description of this 
\emph{equivariant dimensional reduction} can be found in \cite{GP02},
whereas brief reviews of the procedure and the resulting quiver gauge theories
can be found e.g. in \cite{LPS07,DS11}. Given a Hermitian vector bundle $\mathcal{E} \rightarrow M^d \times G/H$ of rank $k$ such that the group $G$ 
acts trivially on $M^d$,
equivariance with respect to $G$ means that the diagram  
\beq
 \xymatrix @=2.5em{\ar @{} [dr] \mathcal{E}\ar[r]^{G \curvearrowright } \ar[d]^{\pi} & \mathcal{E}\ar[d]^{\pi} \\
           M^d\times G/H\ar[r]^{G \curvearrowright} & M^d\times G/H}
\eeq
commutes, and the action of $G$ on the bundle $\mathcal{E}$ induces an
isomorphism between the fibres $\mathcal{E}_x$ and
$\mathcal{E}_{g\cdot x}$ for all 
$x \in M^d \times G/H$. Since the group $H$ acts trivially on the base space, equivariance of the bundle induces a representation of $H$ on the fibres 
$\mathcal{E}_x \simeq \C^k$. Consequently to obtain $G$-equivariant
bundles of a given rank $k$, one has to study (smaller) $H$-representations inside 
the group $\mathrm{U}(k)$ which is the generic structure group of the
bundle acting on the fibres. This representation can be taken to 
descend from the restriction of an irreducible
\mbox{$G$-representation $\mathcal{D}$} comprised of irreducible
$\su$-representations on $\C^{m_1+1}$ and $\C^{m_2+1}$ as $\mathcal{D}|_H = \bigoplus_{i,\a}\, \rho_{i\a}$, which implies that the 
structure group $\mathrm{U}(k)$ is reduced to the subgroup
\beq
\prod_{i=0}^{m_1} \ \prod_{\a=0}^{m_2}
\, \mathrm{U}(k_{i\a}) \ \subset \ \mathrm{U}(k) \ , \ \ \ \ \ \ \ \ \ \ \sum_{i=0}^{m_1} \
\sum_{\a=0}^{m_2} \, k_{i\a}\=k \ .
\label{eq20a}
\eeq
The labelling by \emph{two} indices is due to
the special choice of $G$ here as a product of two Lie groups. Any $G$-equivariant bundle $\mathcal{E} \rightarrow M^d \times G/H$ restricts to an $H$-equivariant bundle $E=\Ecal|_{M^d}$ over $M^d$; inversely, any $H$-equivariant bundle $E\rightarrow M^d$ induces a $G$-equivariant bundle $\mathcal{E}=G \times_{H} E$ over $M^d\times G/H$~\cite{GP02}. 

\medskip

\textbf{Example. \ } To clarify the resulting structures of the
bundles involved and for later comparisons, let us briefly review this construction for the 
K\"ahler manifold $\C P^1 \times \C P^1$ \cite{LPS06}, i.e.~$G=\su \times
\su$ and \mbox{$H=\uo \times \uo$} (instead of $\uo$), which will be 
used as reference in the remainder of this paper. 
The isotopical decomposition of an $H$-equivariant vector bundle 
$E \rightarrow M^d$ of rank $k$ reads~\cite{LPS06}
\beq
E\=  \bigoplus_{i=0}^{m_1} \ \bigoplus_{\a=0}^{m_2}\, {E}_{i\a}
\otimes \underline{S}_{\, m_1-2i}^{(1)} \otimes \underline{S}_{\,
  m_2-2\a}^{(2)} \ ,
\eeq
where one uses a Levi decomposition of the complexified group
$G_{\C}$. The vector spaces $\underline{S}^{(l)}_{\,p_l}\simeq\C$ are 
irreducible $\uo$-representations of weight $p_l$, and the remaining generators of $G$
act non-trivially only on the bundles $E_{i\a}\to M^d$ of rank $k_{i\a}$, 
providing ladder operators due to the $\su$ commutation relations. By induction of bundles, 
a $G$-equivariant bundle $\mathcal{E} \rightarrow M^d \times G/H$ admits the decomposition  
\beq
\mathcal{E} \=  \bigoplus_{i=0}^{m_1} \ \bigoplus_{\a=0}^{m_2} \,
{E}_{i\a} \otimes \mathcal{L}_{(1)}^{m_1-2i}  \otimes
\mathcal{L}_{(2)}^{m_2-2\a} \ , 
\eeq
where 
\beq
\mathcal{L}_{(l)}^{p_l}\=\su \times_{\uo} \underline{S}^{(l)}_{\, p_l}
\label{eq20c}
\eeq
are the monopole line bundles over $\C P^1$ with monopole charge $p_l$. For a rigorous mathematical
treatment of $\su$-equivariant bundles over $\C P^1$
see~\cite{Bismas}.

\medskip

The structural features of the decomposition of $G$-equivariant vector bundles
for a chosen $G$-module $\mathcal{D}$
can be encoded in quivers. For this, one draws a vertex for each irreducible $H$-representation $\rho_{i\a}$
and depicts by arrows the homomorphisms between two representations
$\rho_{i\a}\to \rho_{j\b}$ induced by the action of the entire group $G$.
Thus the restriction of the representation $\mathcal{D}$ leads to an oriented graph
which encodes the field content of the gauge theory. After dimensional reduction of pure Yang-Mills theory on $M^d \times G/H$ to the spacetime $M^d$, 
the arrows of the quiver  constitute a scalar potential for the gauge theory on $M^d$; for this reason 
the corresponding fields are sometimes referred to as $\emph{Higgs
  fields}$, and we shall adapt this nomenclature in the following.


\subsection{Representations of $\, {\rm Spin}(4)$}

Following the approach outlined above, we have to study the irreducible representations of the group $G=\su \times \su$, and we shall start from  
the defining representation of $\su$ on $\C^2$. Since we need in
particular the $H$-representation inside that of the entire structure group, it is convenient to choose the representation by the 
diagonal Pauli matrix and the two ladder operators
\bea
\label{eq21}
\sigma_3 \= \begin{pmatrix}
            1  & 0\\
            0  & -1 
            \end{pmatrix} \ ,  \ \ \ \ \ \ \ \ \ \ \
\sigma_+ \= \begin{pmatrix}
            0  & 1\\
            0  & 0 
            \end{pmatrix} \ ,   \ \ \ \ \ \ \ \ \ \ \
\sigma_- \= \begin{pmatrix}
            0  & 0\\
            1  & 0 
            \end{pmatrix} \ ,
\eea
with the usual commutation relations
\beq
\label{eq23}
\com{\sigma_3}{\sigma_{\pm}}\= \pm\, 2 \sigma_{\pm} \ , \ \ \ \ \ \ \
\com{\sigma_+}{\sigma_-}\= \sigma_3 \ .
\eeq
For any positive integer $m$ one obtains the generalization to an
irreducible representation on $\C^{m+1}$ given by the 
matrices
\bea
\label{eq24}
\nonumber I_{(m)}^+ \= \begin{pmatrix}
             0 & \gamma_0 & 0 & \cdots & 0\\
             0 & 0        & \gamma_1 & \cdots &0\\
             \vdots& \vdots & \vdots & \ddots & \vdots\\
             0  & 0 & 0 & \cdots & \gamma_{m-1}\\
             0  & 0 & 0 & \cdots & 0
             \end{pmatrix} \ , \ \ \  \ \ \ \ I_{(m)}^- \= \big(
             I_{(m)}^+\big)^{\dagger} \ ,\\[4pt]
             I_{(m)}^3 \= \mathrm{diag}\lb m, m-2, \ldots, -m+2, -m\rb
             \ , 
\eea
with $\gamma_j^2 \coloneqq \lb j+1\rb \lb m-j\rb$ for $j=0,1, \ldots, m-1$, yielding the relations (\ref{eq23}). Irreducible representations of 
the group $G$ are then given by the tensor product of two single $\su$-representations on
$\C^{m_1+1}\otimes \C^{m_2+1}$, and the six generators read
\beq
\label{eq28}
I_{(m_1)}^{\pm} \otimes \id_{m_2+1} \ , \ \ I_{(m_1)}^3 \otimes
\id_{m_2+1} \ , \ \ \id_{m_1+1} \otimes I_{(m_2)}^{\pm} \ , \ \
\id_{m_1+1} \otimes I_{(m_2)}^3 \ .
\eeq
Since $T^{1,1}$ is a reductive homogeneous space, one has the splitting
\beq
 \mathfrak{g} \ \coloneqq \ \mathfrak{su}(2) \oplus \mathfrak{su}(2)
 \= \mathfrak{u}(1)\oplus \mathfrak{m} \ =: \ \mathfrak{h} \oplus
 \mathfrak{m}
\label{eq29b}
\eeq
with $[\mathfrak{h}, \mathfrak{m}]\subset \mathfrak{m}$,
where the Lie algebra $\mathfrak{h}$ is generated by the difference of the two diagonal operators, 
and the five-dimensional complement $\mathfrak{m}$ can be identified with the cotangent space 
of $T^{1,1}$. By definition and construction of the basis $\left\{e^{\m}\right\}$ and the monopole form $a$,
the ladder operators on the two copies of $\su$ are dual to the complex forms $e^1 \pm \im e^2$ and $e^3 \pm \im e^4$, while the forms 
$a$ and $\frac{3\im}{4}\, e^5$ correspond to the difference and sum, respectively, of the diagonal operators. 

Since the existence of a $G$-equivariant structure on a vector bundle is accompanied by a
reduction of its structure group according to (\ref{eq20a}),  the
direct sum
of $H$-representations $\C^{k_{i\alpha}} \coloneqq \C^{k_i} \otimes \C^{k_{\alpha}}$, 
\beq
\label{eq31}
\C^{k} \= \bigoplus_{i=0}^{m_1} \ \bigoplus_{\alpha=0}^{m_2}\,
\C^{k_{i\alpha}} \ , \ \ \ \ \ \ \ \ \ k\= \sum_{i=0}^{m_1} \
\sum_{\alpha=0}^{m_2}\, k_{i\alpha} \ ,
\eeq 
must be studied under the action of the group $G$. Due to the block form of the broken structure group and in the spirit of how vertices 
of the quiver  arise, it is convenient to interpret vectors
in the space $\C^k$ as vectors of length $\lb m_1+1\rb \lb m_2+1\rb$ 
whose entries are vectors in the spaces $\C^{k_{i\a}}$ rather than
complex numbers, as dictated in the decomposition \eqref{eq31}. Then each entry of the vector corresponds exactly to one vertex $v_{i\a}$ in the quiver, and arrows 
occur if there is a non-vanishing homomorphism in $\Hom\big(
\C^{k_{i\a}}, \C^{k_{j\b}} \big)$ as an entry in the block matrices which describe the action 
of $G$. The representation of the group action in terms of the
generators given above can be adapted to the vector space $\C^k$  by 
keeping the general form of (\ref{eq24}) and substituting the complex
numbers as entries by matrices. We will mostly assume this 
convention implicitly in the following.  

Finally, for a more convenient description of the generator $I_6$ of
$\mathfrak{h}$ on $\C^k$, we introduce natural projection operators on $\C^{m_1+1}$ and 
$\C^{m_2+1}$, respectively, by \cite{LPS06}
\bea
\label{eq34}
\nonumber &\Pi_i : \C^{m_1+1} \longrightarrow \C \ , \ \ \ \ &\Pi_i \=
\mathrm{diag} ( 0, \ldots,\underset{\stackrel{\uparrow}{\mathrm{i-th\
      slot}}}{0,1,0,} \ldots, 0) \ , \ \ \ i\=0,1, \ldots, m_1 \ ,\\[4pt]
&\Pi_{\a} : \C^{m_2+1} \longrightarrow \C \ , \ \ \ \ &\Pi_{\a} \=
\mathrm{diag} ( 0, \ldots,\underset{\stackrel{\uparrow}{\mathrm{\a-th\
      slot}}}{0,1,0,} \ldots, 0) \ , \ \ \ \alpha \=0,1, \ldots, m_2 \
,
\label{eq_ident}
\eea
where Latin indices always refer to the first copy of $\su$ and Greek indices to the second copy.
The projection from the tensor product $\C^{m_1+1}\otimes \C^{m_2+1}$ to the component with indices $i$ and $\a$ is thus given by the operator
\beq
\label{eq35}
\Pi_{i\alpha} : \C^{m_1+1}\otimes \C^{m_2+1} \longrightarrow \C \ , \ \ \ \ \Pi_{i\a} \ \coloneqq \ \Pi_i \otimes \Pi_{\a}
\eeq
and thus by the diagonal square matrix 
\beq
\Pi_{i\alpha}\= \big( \delta_{ij}\, \delta_{\a\b}\, \delta_{ik}\, \delta_{\a\g}\big)
_{\b,\g=0,1, \ldots, m_2}^{j,k=0,1, \ldots, m_1}
\label{eq:proj}
\eeq
of size $\left[\lb m_1+1\rb \lb m_2+1\rb\right]^2$.
Furthermore, we introduce the operators
\bea
\label{eq36}
\nonumber \Pi_i^{(1)} &\coloneqq & \sum_{\a=0}^{m_2}\, \Pi_{i\a} \=
\Pi_i \otimes \sum_{\a=0}^{m_2}\, \Pi_{\a} \= \Pi_i \otimes
\id_{m_2+1} \ , \\[4pt]
\Pi_{\a}^{(2)} &\coloneqq & \sum_{i=0}^{m_1}\, \Pi_{i\a} \=
\sum_{i=0}^{m_1}\, \Pi_i \otimes \Pi_{\a} \= \id_{m_1+1} \otimes
\Pi_{\a}  \ ,
\eea
which project on all components with a fixed value of the first or second index, respectively. 
Interpreting (implicitly) all entries $1$ as the identity operator $\id$ of the pertinent dimension, one obtains a representation of the generators of the 
maximal torus of $\su \times \su$ on the vector space \eqref{eq31} by the diagonal matrices
\bea
\label{eq37}
\nonumber  \Upsilon^{(1)} &\coloneqq & \sum_{i=0}^{m_1}
\, \lb m_1-2i \rb \Pi_i^{(1)} \= I_{(m_1)}^3 \otimes \id_{m_2+1} \ ,\\[4pt]
\Upsilon^{(2)} &\coloneqq & \sum_{\a=0}^{m_2} \, \lb
m_2-2\a\rb \Pi_{\a}^{(2)} \= \id_{m_1+1} \otimes I_{(m_2)}^3 \ . 
\eea
In particular, the Lie algebra $\mathfrak{h}$ is generated by 
\bea
 I_{6} \= \Upsilon^{(1)}-\Upsilon^{(2)}
                            \= \sum_{i=0}^{m_1} \
                            \sum_{\alpha=0}^{m_2} \, \lb m_1-m_2 -2 i
                            + 2\alpha\rb \Pi_{i\alpha} \ .
\label{eq38}
\eea


\subsection{Representations of quivers}
Based on the previous algebraic description of the generators of $G$ and their
representation on the vector space $\C^k$, the forms of the Higgs fields and the
quivers can already be deduced to some extent, but to actually construct the most general connection that is compatible with
the equivariance of the bundles we have to formulate the construction more precisely. For this, we start from the canonical connection 
$\Gamma= I_6 \otimes a$ and recall that the space
$T^{1,1}$ is reductive, i.e. one has the commutation relations
\beq
\com{I_6}{I_{\m}} \= f_{6\m}^{\n}\, I_{\n} \ , \ \ \ \ \ \ \ \ \  \ \
\ \ \com{I_{\m}}{I_{\n}} \= f_{\m\n}^6\, I_6 + f_{\m\n}^{\rho}\,
I_{\rho} \ , \ \ \ \ \ \ \ \m,\n,\rho \=1,
\ldots, 5 \ ,
\label{eq38b}
\eeq  
according to the decomposition of $\mathfrak{g}$ into the Lie algebra
$\mathfrak{h}$ of $\uo$ and its complement $\mathfrak{m}$. A
connection $\mathcal{A}$ on the $G$-equivariant bundle $\mathcal{E}\to M^d \times T^{1,1}$ can 
generically be written as 
\bea
\label{eq39}
 \ca          &=& A + \Gamma + X_{\mu}\otimes e^{\mu}\\[4pt]
\nonumber     &=:& A + \Gamma +\hf^{(1)}
\otimes\lb e^1-\im e^2\rb -\hf^{(1)}\,^\dagger \otimes\lb e^1+\im e^2\rb+ \hf^{(2)} \otimes\lb e^3-\im e^4\rb -
\hf^{(2)}\,^\dagger \otimes\lb e^3+\im e^4\rb \\ \nonumber && +\, \hf^{(3)} \otimes e^5 \
,
\eea
where $A$ is a connection on the corresponding $H$-equivariant bundle $E\to M^d$, and
where we have combined the bundle endomorphisms, expressed by the 
skew-Hermitian matrices $X_{\mu}$, into the quantities 
\beq
\phi^{(1)} \ \coloneqq \ \mbox{$\frac{1}{2}$} \lb X_1+\im X_2\rb \ , \ \ \ \
\phi^{(2)} \ \coloneqq \ \mbox{$\frac{1}{2}$} \lb X_3 +\im X_4\rb \qquad\textrm{and}\qquad
\phi^{(3)} \ \coloneqq \ X_5 \ ,
\eeq
which we call Higgs fields. 
The Higgs fields $\phi^{(1)}$ and $\phi^{(2)}$ accompany the anti-holomorphic 
\mbox{1-forms}\footnote{
The action of the complex structure $J$ induced by the contact
structure on the leaf spaces
is given by $Je^1 =-e^2$, $Je^2=e^1$, $Je^3=-e^4$, and $Je^4=e^3$. The
corresponding K\"ahler form is given by 
$\omega \lb \cdot, \cdot\rb \coloneqq g \lb J \cdot, \cdot\rb= e^{12}+e^{34}= \omega^3$.} $\bar{\Theta}{}^1 \coloneqq e^1-\im e^2$ and 
$\bar{\Theta}{}^2 \coloneqq  e^3-\im e^4$. 
The tensor product symbol between the endomorphism part and the form part will be omitted from now on.

For the connection (\ref{eq39}) to be compatible with the equivariance of the underlying bundle, we have to impose two conditions that can be directly 
gleamed from the structure of the field strength. Setting
$A=0$ for the time being, the curvature of the connection reads
\bea
 \nonumber    \cf&=&\diff \ca +\ca \wedge \ca \\[4pt]
\nonumber      &=& \big( \com{I_6}{X_1}-2\im X_2\big)\, a\wedge e^1 + \big( \com{I_6}{X_2}+2\im X_1\big)\, a\wedge e^2+ \big( \com{I_6}{X_3}+2\im X_4\big)\, a\wedge e^3\\
\nonumber     && + \, \big( \com{I_6}{X_4}-2\im X_3\big)\, a\wedge
e^4+\com{I_6}{X_5} a \wedge e^5\\
\nonumber     && +\, \big(\com{X_1}{X_2}-2X_5+\mbox{$\frac{3}{2}$}\,
\im I_6\big)\, e^{12}+\big(
\com{X_3}{X_4}-2X_5-\mbox{$\frac{3}{2}$}\, \im I_6\big)\, e^{34}+\big(
\com{X_1}{X_5}+\mbox{$\frac{3}{2}$}\, X_2\big)\, e^{15}\\ 
\nonumber     && +\,\big( \com{X_2}{X_5}-\mbox{$\frac{3}{2}$}\,
X_1\big)\, e^{25} +\big( \com{X_3}{X_5}+\mbox{$\frac{3}{2}$}\,
X_4\big)\, e^{35}+\big( \com{X_4}{X_5}-\mbox{$\frac{3}{2}$}\, X_3\big)
\, e^{45}\\  
\label{eq41} && +\phantom{(}\com{X_1}{X_3} \, e^{13}+\com{X_1}{X_4}
\, e^{14} +\com{X_2}{X_3} \, e^{23} +\com{X_2}{X_4} \, e^{24}+ \diff
X_{\m} \wedge e^{\m} \ .              
\eea
The $G$-equivariance is spoiled by terms involving a mixture of \mbox{1-forms} in  $\mathfrak{g}^*$ and $\mathfrak{h}^*$, i.e. by the occurence of
\mbox{2-forms} $a\wedge e^{\m}$. Therefore, firstly, one assumes that the Higgs fields do not depend on the coordinates of the coset space but only on those 
of the spacetime $M^d$, which ensures that the sum in the very last term does not contain incompatible \mbox{2-forms}. 
Moreover, as an additional benefit, this condition greatly simplifies the dimensional reduction of the gauge theory.

Secondly, the first five terms in (\ref{eq41}) must vanish, and this
requirement determines the features of the quiver gauge theory.
Supposing that the sum of the $H$-representations stems from an irreducible representation of $G$, for compatibility one may demand that the 
endomorphisms $X_{\m}$ act in the same way
on the fibres $\C^k$ as the generators (\ref{eq38b}) of the Lie algebra $\mathfrak{g}$ do, i.e.
\beq
\com{I_6}{X_{\m}} \= f_{6\m}^{\n}\, X_{\n} \ 
.
\label{eq_cond}
\eeq
Imposing this \emph{equivariance condition} on the Higgs fields is equivalent to the vanishing of the terms that could spoil the
equivariance. This condition can also be motivated \cite{Lubbe} by recalling that a $G$-invariant connection of a vector bundle
over a reductive homogeneous space can be parameterized~\cite{KN} by linear maps $\Lambda: \mathfrak{g}\rightarrow \mathfrak{m}$ such that
\beq
\Lambda\big( \com{W}{ Y} \big) \= \com{W}{ \Lambda (Y)}
\eeq
for all $W \in \mathfrak{h}$ and $Y \in \mathfrak{m}$. By putting
$X_\m=\Lambda( I_\m)$, one directly obtains the conditions
(\ref{eq_cond}) from the relations \eqref{eq38b}. It
forces the underlying graph of the quiver to coincide with the weight diagram of the given representation of $\mathfrak{g}$ if $\mathfrak{h}$ is the Cartan
subalgebra. Enlarging the subspace $\mathfrak{m}$ leads to fewer
restrictions among the Higgs fields. 
For the  Higgs fields associated to $T^{1,1}$ one obtains 
the necessary conditions
\bea\label{eq42}
\com{\Upsilon^{(1)}-\Upsilon^{(2)}
}{\hf^{(1)}} \= 2 \hf^{(1)} \ , \quad
\com{\Upsilon^{(1)}-\Upsilon^{(2)}
}{\hf^{(2)}} \= -2 \hf^{(2)} \ ,\quad
\com{\Upsilon^{(1)}-\Upsilon^{(2)} }{
  \hf^{(3)}} \= 0 \ .
\eea

To evaluate these relations explicitly, let $E_{i \alpha, j \beta}$ be
the square matrix of size $\left[(m_1+1)\, (m_2+1)\right]^2$ with the entry 
$1$ (again interpreted as an identity operator) at the position $(i\alpha, j\beta)$ and zero otherwise, yielding the commutation relations
\beq
\label{eq43}
\com{E_{i\alpha,j\beta}}{E_{k\gamma,l\delta}} \= \delta_{jk}\,
\delta_{\beta\gamma}\, E_{i\alpha, l\delta} -\delta_{il}\,
\delta_{\alpha\delta}\, E_{k\gamma, j\beta} \ ,
\eeq
and in particular
\beq
\label{eq44}
\com{\Pi_{i\alpha}}{E_{j\beta, k\gamma}} \= \delta_{ij}\,
\delta_{\alpha\beta} \, E_{i\alpha, k\gamma}- \delta_{ik}\,
\delta_{\alpha\gamma} \, E_{j\beta, i \alpha} \ .
\eeq
We decompose the Higgs fields according to their block structure as 
\beq
\hf^{(a)}\= \sum_{j,k=0}^{m_1} \ \sum_{\beta,\gamma=0}^{m_2}\, \phi^{(a)}_{j\beta, k\gamma} \, E_{j\beta, k\gamma} \qquad\textrm{for}\quad a\=1,2,3 \ ,
\eeq
where $\phi^{(a)}_{j\beta, k\gamma}\in {\rm Hom}(E_{k\gamma},E_{j\beta})$. Then using (\ref{eq44}) the commutators read
\bea
\com{\Upsilon^{(1)}-\Upsilon^{(2)}}{\hf^{(a)}} 
\= \sum_{j,k=0}^{m_1} \ \sum_{\beta,\gamma=0}^{m_2}\, 2 \lb k-j-
\gamma +\beta \rb \phi^{(a)}_{j\beta, k\gamma} \, E_{j\beta, k\gamma}
\ .
\label{eq45}
\eea
Hence the conditions (\ref{eq42}) restrict the component homomorphisms of the Higgs fields to be of the form
\bea
\nonumber  \phi^{(1)}_{j\beta, k\gamma} &=&
\delta_{k-j-\gamma+\beta,1}\ \phi^{(1)}_{j\beta, k\gamma} \ , \qquad
 \phi^{(2)}_{j\beta, k\gamma} \= \delta_{k-j-\gamma +\beta,-1}\
 \phi^{(2)}_{j\beta, k\gamma} \ ,\\[4pt]
\phi^{(3)}_{j\beta, k\gamma} &=& \delta_{k-j-\gamma +\beta,0}\
\phi^{(3)}_{j\beta, k\gamma} \ .
\label{eq46}
\eea
This implies that $\phi^{(1)}$ and $\phi^{(2)}$ act as ladder
operators which increase or decrease, respectively, by one unit
the \emph{relative} quantum number which is the difference of monopole
charges at the vertices of the quiver given by
\beq 
c_{i\a} \ \coloneqq \ m_1-m_2-2i +2\a \ .
\eeq
In particular, we cannot associate the action of any one of 
these Higgs fields to only a single copy of $\su$ in the tensor product as
in the case of the quiver gauge theories associated to $\C P^1\times
\C P^1$, as here they act 
simultaneously on both components. Since we allow for arbitrary entries of the Higgs fields in quiver gauge theory, 
there are generically \emph{two} arrows (with opposite orientations) between vertices whose relative quantum numbers $c_{i\a}$ differ by one unit.
Note that $\phi^{(1)}$ is not related to the adjoint
bundle morphism of the field $\phi^{(2)}$, and so the pertinent quiver
is generically the \emph{double} of an underlying quiver; in this
sense the resulting quiver gauge theories are analogous to those
obtained via dimensional reduction over quasi-K\"ahler coset spaces~\cite{Popov:2010rf}. On the other hand, the 
endomorphism $\phi^{(3)}$ represents the contribution from the
vertical components of the Sasakian fibration and yields arrows conserving 
the relative quantum number $c_{i\a}$. This induces a loop at each vertex as well as arrows between vertices carrying the same $c_{i\a}$ value,
which realize different partitions of a given difference of the indices $i$ and $\a$. This less restrictive property of the Higgs fields is 
caused by factoring out a smaller subalgebra
$\mathfrak{h}$. Nonetheless, the equivariance conditions still rule out many possible arrows.
 

\subsection{Examples}
The general form of the Higgs fields and the quivers obtained by imposing $G$-equivariance over $T^{1,1}$ are completely 
dictated by the conditions (\ref{eq46}). In order to gain a better
insight into the structures obtained from these relations, we shall
consider three explicit examples.

\medskip

$\mbf{(m_1,m_2)=(m,0).} \ $ In this case the representation labelled by the second index acts trivially and the tensor product reduces to 
a representation of the first $\su$ factor. The conditions (\ref{eq46}) require the Higgs fields $\phi^{(1)}$ and $\phi^{(2)}$
to connect adjacent vertices (acting in opposite directions), while $\phi^{(3)}$ creates a loop at each vertex. 
Therefore this choice of representation yields the double of the $A_{m}$ quiver~\cite{LPS06,SL2C,PS06} with vertex loops
\beq
    \begin{tikzpicture}[->,scale=3]
    \node (a) at (0,0) {$\mbf{(0,0)}$};
    \node (b) at (1,0) {$\mbf{(1,0)}$};
    \node (c) at (2,0) {$ \ \ldots \ $};
    \node (d) at (3,0) {$\mbf{(m-1,0)}$};
    \node (e) at (4,0) {$\mbf{(m,0)}$};

    \path[thick](b) edge [bend left=15] node {} (a);
    \path[thick](c) edge [bend left=15] node {} (b);
    \path[thick](d) edge [bend left=15] node {} (c);
    \path[thick](e) edge [bend left=15] node {} (d);
     \path[thick,->, dashed,blue] (a) edge [bend right=-15] node {} (b);
     \path[thick,->, dashed,blue] (b) edge [bend right=-15] node {} (c);
     \path[thick,->, dashed,blue] (c) edge [bend right=-15] node {} (d);
     \path[thick,->, dashed,blue] (d) edge [bend right=-15] node {} (e);
    \draw [thick,->, dotted, thick,red](a) to [out=70,in=110,looseness=13] (a);
    \draw [thick,->, dotted, thick,red](b) to [out=70,in=110,looseness=13] (b);
    \draw [thick,->, dotted, thick,red](d) to [out=70,in=110,looseness=13] (d);
    \draw [thick,->, dotted, thick,red](e) to [out=70,in=110,looseness=13] (e);
\end{tikzpicture}
\eeq
where the vertices are labelled by their indices $(i,\a)$,
and the arrows represent $\phi^{(1)}$ (solid lines), $\phi^{(2)}$~(dashed lines), and $\phi^{(3)}$ (dotted lines). For the equivariant connection in the simplest case $m=1$ one obtains the explicit form
\bea
\ca \= \begin{pmatrix}
                a +\Phi^{(3)}_{0,0} & \Phi^{(1)}_{0,1}- \Phi^{(2)\, \dagger}_{1,0}\\
                -\Phi^{(1)\, \dagger}_{0,1} + \Phi^{(2)}_{1,0} & -a + \Phi^{(3)}_{1,1}
               \end{pmatrix} \ ,
\eea
where we have defined 
\beq
\Phi^{(1)}_{i,j} \ \coloneqq \ \phi^{(1)}_{i,j} \lb e^1-\im
e^2\rb \ , \ \ \ \ 
\Phi^{(2)}_{i,j} \ \coloneqq \ \phi^{(2)}_{i,j} \lb e^3-\im e^4\rb \qquad\textrm{and}\qquad
\Phi^{(3)}_{i,j} \ \coloneqq \ \phi^{(3)}_{i,j} \, e^5 \ , 
\eeq
omitting Greek indices which refer to the second trivial factor of the tensor product. 
We notice that even for this simple case, by the weaker conditions on
the Higgs fields, there are \emph{two} contributions to the off-diagonal
components of the connection. 

\medskip

$\mbf{(m_1,m_2)=(1,1). \ }$ In this representation the $\mathrm{U}(1)$ generator $I_6$ and its commutator 
with an arbitrary $4\times4$ matrix $\lb \bullet \rb$ are given by 
\bea
I_6 \= \begin{pmatrix} 
        0 & 0& 0&0\\
        0 & 2 & 0 & 0\\
        0 & 0& -2 & 0\\
        0 & 0& 0 & 0 
        \end{pmatrix} \ , \ \ \ \
\com{I_6}{(\bullet)} \=  \begin{pmatrix}
    \phantom{-}0\bullet & -2\bullet& \phantom{-}2\bullet& \phantom{-}0\bullet\\
    \phantom{-}2\bullet& \phantom{-}0\bullet& \phantom{-}4\bullet& \phantom{-}2\bullet\\
    -2\bullet& -4\bullet& \phantom{-}0\bullet& -2\bullet\\
    \phantom{-}0\bullet& -2\bullet& \phantom{-}2\bullet& \phantom{-}0\bullet
   \end{pmatrix} \ ,
\eea
which forces, according to (\ref{eq42}), the Higgs fields to be of the form
\bea
\phi^{(1)} \= \begin{pmatrix}
                   0 & 0& \ast & 0\\
                   \ast & 0 & 0 & \ast\\
                    0 & 0 & 0 & 0\\
                    0 & 0 & \ast & 0
                 \end{pmatrix} \ ,\ \ \
\phi^{(2)}\= \begin{pmatrix}
                     0 & \ast & 0 &0\\
                     0 &0 & 0 &0\\
                     \ast& 0 & 0 &\ast\\
                     0 & \ast & 0 & 0  
                     \end{pmatrix} \ , \ \
 \phi^{(3)} \= \begin{pmatrix}
              \ast & 0 &0 &\ast\\
               0   & \ast & 0 &0\\
               0   & 0    & \ast & 0\\
               \ast   & 0 &  0 & \ast
             \end{pmatrix} \ ,
\eea
in accordance with (\ref{eq46}). The corresponding quiver
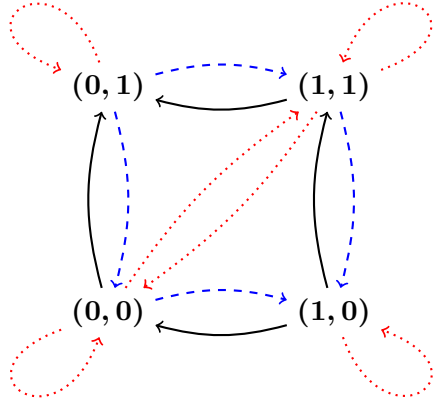
\beq
    \begin{tikzpicture}[->,scale=3]
        \node (a) at (0,0) {$\mbf{(0,0)}$};
        \node (b) at (0,1) {$\mbf{(0,1)}$};
        \node (c) at (1,0) {$\mbf{(1,0)}$};
        \node (d) at (1,1) {$\mbf{(1,1)}$};

        \path[thick, ->] (c) edge [bend left=15]  node  {} (a);
         \path[thick, ->] (a) edge [bend left=15]  node  {} (b);
          \path[thick, ->] (d) edge [bend left=15]  node  {} (b);
          \path[thick, ->] (c) edge [bend left=15]  node  {} (d);

         \path[thick,->, dashed, blue] (b) edge [bend right=-15] node {} (a);
         \path[thick,->, dashed, blue] (a) edge [bend right=-15] node {} (c);
         \path[thick,->, dashed, blue] (d) edge [bend right=-15] node {} (c);
         \path[thick,->, dashed, blue] (b) edge [bend right=-15] node {} (d);

         \draw [thick,->, dotted, thick, red](a) to [out=200,in=250,looseness=8] (a);
        \draw [thick,->, dotted, thick, red](b) to [out=110,in=160,looseness=8] (b);
        \draw [thick,->, dotted, thick, red](c) to [out=290,in=340,looseness=8] (c);
        \draw [thick,->, dotted, thick, red](d) to [out=20,in=70,looseness=8] (d);
        \path[thick,->, dotted, thick, red] (d) edge [bend left=10] node {} (a);
        \path[thick,->, dotted, thick, red] (a) edge [bend right=-10] node {} (d);

\end{tikzpicture}
\eeq
is a square lattice with double arrows as underlying graph and, additionally, 
the vertex loop modifications we have already encountered in the previous example. However, there is also a further arrow induced by $\phi^{(3)}$ because 
the vertices $(0,0)$ and $(1,1)$ realize the same relative quantum number $c_{i\a}=2 \lb \a -i \rb$.
The compatible connection then reads 
\bea
\ca \= \begin{pmatrix}
         \Phi^{(3)}_{00,00}  &  -\Psi_{01,00}{}^{\dagger}  & \Psi_{00,10} &  \Phi^{(3)}_{00,11} \\
        \Psi_{01,00}  & 2a+\Phi^{(3)}_{01,01}  & 0 & \Psi_{01,11}\\
        -\Psi_{00,10}{}^{\dagger} & 0 & -2a+ \Phi^{(3)}_{10,10} & -\Psi_{11,10}{}^{\dagger}\\
        - \Phi^{(3)}_{00,11}{}^\dagger  & -\Psi_{01,11}{}^{\dagger}& \Psi_{11,10}& \Phi^{(3)}_{11,11}
       \end{pmatrix} \ ,
\eea
where we have set $\Psi_{i\a,j\b}\coloneqq
\Phi^{(1)}_{i\a,j\b}-\Phi^{(2)}_{i\a,j\b}{}^\dagger$, using again the
definitions introduced above. 

\medskip

$\mbf{(m_1,m_2)=(2,1).} \ $ With the representation $I_6=\mathrm{diag}(1,3,-1,1,-3,-1)$ 
one obtains the commutator
 \beq
\com{I_6}{(\bullet)} \= \begin{pmatrix}
                          \phantom{-}0\bullet &  -2\bullet& \phantom{-}2\bullet& \phantom{-}0\bullet& \phantom{-}4\bullet& \phantom{-}2\bullet\\
                          \phantom{-}2\bullet &  \phantom{-}0\bullet& \phantom{-}4\bullet& \phantom{-}2\bullet& \phantom{-}6\bullet& \phantom{-}4\bullet \\
                          -2\bullet &  -4\bullet& \phantom{-}0\bullet& -2\bullet& \phantom{-}2\bullet& \phantom{-}0\bullet \\
                           \phantom{-}0\bullet &  -2\bullet& \phantom{-}2\bullet& \phantom{-}0\bullet& \phantom{-}4\bullet& \phantom{-}2\bullet \\
			   -4\bullet &  -6\bullet& -2\bullet& -4\bullet& \phantom{-}0\bullet& {-}2\bullet \\
                         -2\bullet &  -4\bullet& \phantom{-}0\bullet& -2\bullet& \phantom{-}2\bullet& \phantom{-}0\bullet 
                          \end{pmatrix} \ ,
\eeq
which leads to endomorphisms of the form
\small
\beq
\phi^{(1)} = \begin{pmatrix}
                  0 & 0 & \ast &0     & 0   & \ast   \\
                  \ast & 0    & 0    &\ast  & 0 & 0   \\
                  0 & 0    & 0    & 0  & \ast   & 0\\
                  0 & 0    & \ast    & 0    & 0    & \ast\\
                  0 & 0    & 0    & 0    & 0   & 0\\
                  0 & 0    & 0    & 0    & \ast   & 0
                 \end{pmatrix} \ , \ \ 
\phi^{(2)} = \begin{pmatrix}
                     0		&\ast	&0	&0	&0	&0\\
                     0  	&0	&0	&0	&0	&0\\
                     \ast 	&0	&0	&\ast	&0	&0\\
                     0		&\ast	&0	&0	&0	&0\\
                     0		&0	&\ast	&0	&0	&\ast\\
                     \ast	&0	&0	&\ast	&0	&0
                     \end{pmatrix} \ , \ \
\phi^{(3)} = \begin{pmatrix}
              \ast & 0   & 0   & \ast    & 0    & 0\\
               0   & \ast& 0 &0  & 0    & 0\\
               0   &  0 &\ast & 0    & 0    & \ast\\
               \ast &   0 & 0   & \ast & 0 & 0\\
               0   &   0 & 0   & 0& \ast & 0\\
               0   &   0 & \ast   & 0    & 0    &\ast
             \end{pmatrix} \ .
\eeq
\normalsize
The corresponding quiver
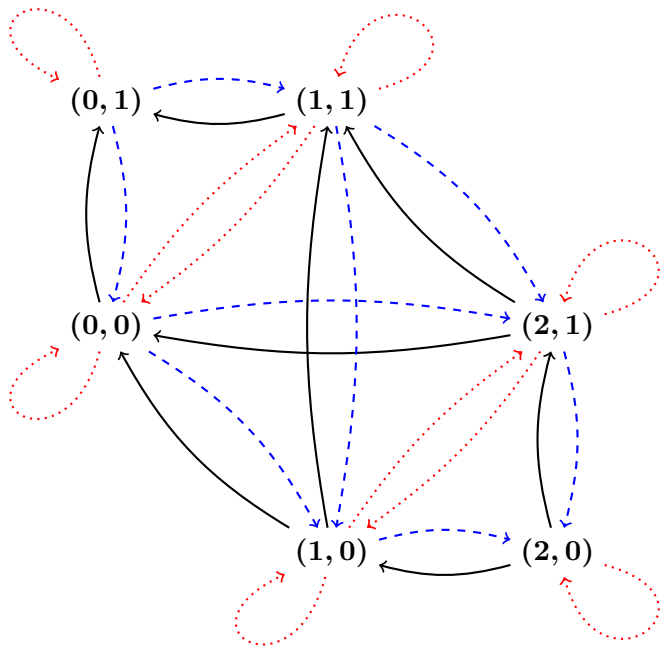
\beq
    \begin{tikzpicture}[->,scale=3]
        \node (a) at (0,1) {$\mbf{(0,0)}$};
        \node (b) at (0,2) {$\mbf{(0,1)}$};
        \node (c) at (1,0) {$\mbf{(1,0)}$};
        \node (d) at (1,2) {$\mbf{(1,1)}$};
        \node (e) at (2,0) {$\mbf{(2,0)}$};
        \node (f) at (2,1) {$\mbf{(2,1)}$};
       
        \draw [thick,->, dotted, thick, red](a) to [out=-105,in=205,looseness=8] (a);
         \draw [thick,->, dotted, thick, red](b) to [out=105,in=-205,looseness=8] (b);
           \draw [thick,->, dotted, thick, red](c) to [out=-105,in=205,looseness=8] (c);
         \draw [thick,->, dotted, thick, red](d) to [out=15,in=75,looseness=8] (d);
          \draw [thick,->, dotted, thick, red](e) to [out=-15,in=-75,looseness=8] (e);
         \draw [thick,->, dotted, thick, red](f) to [out=15,in=75,looseness=8] (f);

        \path[thick,->, dotted, thick, red] (d) edge [bend left=10] node {} (a);
        \path[thick,->, dotted, thick, red] (a) edge [bend right=-10] node {} (d);
        \path[thick,->, dotted, thick, red] (f) edge [bend left=10] node {} (c);
        \path[thick,->, dotted, thick, red] (c) edge [bend right=-10] node {} (f);
        
        \path[thick,->] (c) edge [bend left=15]  node  {} (a);
        \path[thick,->] (f) edge [bend left=10]  node  {} (a);
        \path[thick,->] (a) edge [bend left=15]  node  {} (b);
        \path[thick,->] (d) edge [bend left=15]  node  {} (b);
        \path[thick,->] (e) edge [bend left=15]  node  {} (c);
        \path[thick,->] (c) edge [bend left=10]  node  {} (d);
        \path[thick,->] (f) edge [bend left=15]  node  {} (d);
        \path[thick,->] (e) edge [bend left=15]  node  {} (f);
        \path[thick,->, dashed, blue] (a) edge [bend right=-15] node {} (c);
        \path[thick,->, dashed, blue] (a) edge [bend right=-10] node {} (f);
        \path[thick,->, dashed, blue] (b) edge [bend right=-15] node {} (a);
        \path[thick,->, dashed, blue] (b) edge [bend right=-15] node {} (d);
        \path[thick,->, dashed, blue] (c) edge [bend right=-15] node {} (e);
        \path[thick,->, dashed, blue] (d) edge [bend right=-10] node {} (c);
        \path[thick,->, dashed, blue] (d) edge [bend right=-15] node {} (f);
        \path[thick,->, dashed, blue] (f) edge [bend right=-15] node {} (e);
       
\end{tikzpicture}
\eeq
shows that, besides the additional arrows
between vertices of the same relative quantum number, new arrows
between the vertices $(0,0)$ and $(2,1)$ occur.  

\medskip

Given that the equivariance conditions (\ref{eq46}) and, consequently, the structure of the quivers depend only 
on the relative quantum number~$c_{i\a}$ rather than on the two $\mathrm{U}(1)$ charges separately,
one may combine vertices with the same relative quantum
number and define the homomorphisms between them by appropriate combinations of
the Higgs fields. 
This repackaging implies that the quiver parameterized by $(m_1,m_2)$
is reconsidered as an $(m_1+m_2,0)$ quiver by an equivalence relation on the vertices 
$(i,\alpha)\sim (i+\delta,\alpha+\delta)$ with integral~$\delta$, 
as long as the first entries remain in the interval $[0,m_1]$ and the second entries in $[0,m_2]$.
Geometrically, this means that we project along lines with unit slope in the rectangular graph of the quiver. 
This turns all $\phi^{(3)}$ arrows into vertex loops. 
Hence the quiver gauge theory associated to $T^{1,1}$ for the
decomposition $(m_1, m_2)$ may be interpreted as that of the double of
an $A_{m_1+m_2}$ quiver with one loop at each vertex, in terms of
combinations of fields (with suitable multiplicities). 
This projection is comparable to the collapsing method applied 
for obtaining $\mathrm{SU}(3)$-equivariant quiver gauge theories \cite{SU3,LPS07} starting from weight diagrams of $\mathrm{SU}(3)$.

\subsection{Reduction to $A_{m_1}\oplus A_{m_2}$ quiver gauge theory}

Let us now briefly pause to compare the quiver gauge theory obtained for the
internal manifold $T^{1,1}$ with the quiver gauge theory associated to $\C P^1\times \C P^1$, which is included as a special case in our framework. Starting from scratch, 
one may consider the equivariant dimensional reduction for the splitting $\mathfrak{g}=\mathfrak{m}\oplus \lb \mathfrak{u}(1)\oplus \mathfrak{u}(1)\rb$ by choosing, 
for instance, $\hat{I}_5 = \Upsilon^{(1)}$ and $\hat{I}_6=\Upsilon^{(2)}$ as generators of the subalgebra $\mathfrak{h}=\mathfrak{u}(1)\oplus\mathfrak{u}(1)$, see~(\ref{eq37}). 
Then the resulting equivariance conditions are
\bea
 \com{\Upsilon^{(1)}}{ \phi^{(1)}} \= 2
 \phi^{(1)} \ , \qquad \com{
   \Upsilon^{(2)} }{ \phi^{(1)}} \= 0 \=
 \com{\Upsilon^{(1)} }{ \phi^{(2)}} \ ,
 \qquad \com{\Upsilon^{(2)} }{
   \phi^{(2)}} \= 2 \phi^{(2)} \ ,
\label{eq:cp1cp1_cond}
\eea
and they uniquely determine the Higgs fields to be the ladder
operators of the individual copies of $\su$, recovering the correct
result of~\cite{LPS06}; in particular we recover the rectangular
lattice of the $A_{m_1}\oplus A_{m_2}$ quiver
\beq
    \begin{tikzpicture}[->,scale=2]
        \node (a) at (0,0) {$\mbf{(0,0)}$};
        \node (b) at (1,0) {$\mbf{(1,0)}$};
        \node (c) at (0,1) {$\mbf{(0,1)}$};
        \node (d) at (1,1) {$\mbf{(1,1)}$};
        \node (e) at (2,0) {$\mbf{(2,0)}$};
        \node (f) at (2,1) {$\mbf{(2,1)}$};
        \node (g) at (3,0) { \ $\ldots$};
        \node (h) at (3,1) { \ $\ldots$};
        \node (i) at (0,2) {$\vdots$};
        \node (j) at (1,2) {$\vdots$};
        \node (k) at (2,2) {$\vdots$};

        \path[thick, ->] (b) edge node  {} (a);
        \path[thick,->, dashed, blue] (c) edge  node {} (a);
        \path[thick,->] (e) edge node  {} (b);
        \path[thick,->, dashed, blue] (d) edge  node {} (b);
        \path[thick,->] (g) edge node  {} (e);
        \path[thick,->, dashed, blue] (f) edge  node {} (e);
        \path[thick,->] (h) edge node  {} (f);
        \path[thick,->] (d) edge node  {} (c);
        \path[thick,->] (f) edge node  {} (d);
        \path[thick,->, dashed, blue] (i) edge  node {} (c);
        \path[thick,->, dashed, blue] (j) edge  node {} (d);
        \path[thick,->, dashed, blue] (k) edge  node {} (f);
\end{tikzpicture}
\eeq
generated by $\phi^{(1)}$ (solid lines) and $\phi^{(2)}$ (dashed lines), where the 
vertices are labelled by their indices $\lb i,\a\rb$. The rectangular weight diagram (without double arrows) emerges
here because $H=\uo \times \uo$ is the maximal torus of $G$ in this case.

On the other hand, starting from our previous construction, the $\C P^1 \times \C P^1$ conditions are included in the more general $T^{1,1}$ framework. To recover this special case, 
we have to identify the second generator of $\mathfrak{h}$ in
(\ref{eq39}) and (\ref{eq41}). Taking $X_5$ proportional to the
generator $I_5 =\Upsilon^{(1)}+\Upsilon^{(2)}$, one has to interpret 
$e^5$ as the monopole field strength corresponding to the second generator and to demand a vanishing of the mixed terms, i.e. those containing the forms 
$e^{\m}\wedge e^5$ in (\ref{eq41}), which yields the additional equivariance conditions
\bea 
\com{X_5}{X_1}\= \mbox{$\frac{3}{2}$}\, X_2 \ , \ \ \ \com{X_5}{X_2}\=
-\mbox{$\frac{3}{2}$}\, X_1 \ , \ \ \ \com{X_5}{X_3}\=
\mbox{$\frac{3}{2}$}\, X_4 \ , \ \ \
\com{X_5}{X_4}\=-\mbox{$\frac{3}{2}$}\, X_3 \ .
\label{eq_correspondence}
\eea    
A comparison with the structure constants in (\ref{eq:f_const}) (see Appendix~A) indicates that for the limit one should set 
\mbox{$X_5 =- \frac{3\im}{4}\, \big( \Upsilon^{(1)}+\Upsilon^{(2)}\big)$}, so that 
the further conditions read
\bea
\com{\Upsilon^{(1)}+\Upsilon^{(2)} }{
  \hf^{(1)}}\= 2 \hf^{(1)} \ , \qquad
\com{\Upsilon^{(1)}+\Upsilon^{(2)} }{
  \hf^{(2)}}\= 2 \hf^{(2)} \ .
\eea
Together with (\ref{eq42}), we obtain again the defining relations for the ladder operators of the two $\su$ Lie algebras. 
Thus the quiver gauge theory associated to $\C P^1 \times \C P^1$ (for the correct values of the radii $R_l$) is 
contained in the quiver gauge theory associated to $T^{1,1}$ by taking
the limit $X_5= -\frac{3\im}{4} \big(
\Upsilon^{(1)}+\Upsilon^{(2)}\big)$.

\subsection{Dimensional reduction of Yang-Mills theory}
The $G$-equivariance constraints have strongly restricted the form of
compatible gauge connections $\ca$ on the bundle \mbox{$\Ecal\to M^d \times T^{1,1}$}. We
shall now study the action functional for pure Yang-Mills theory on the product manifold
$M^d \times T^{1,1}$ and then perform the dimensional reduction to an
effective quiver gauge theory on $M^d$, which is a Yang-Mills-Higgs theory with the internal manifold
 providing the non-trivial contributions to the Higgs potential.

After implementation of the equivariance conditions, the non-vanishing components of the field strength 
$\cf=\frac{1}{2} \, \cf_{\hat{\m}\hat{\n}}\, e^{\hat{\m}} \wedge
e^{\hat{\n}}$ read\footnote{
Here indices $a,b, \ldots$ 
refer to $M^d$ and $\m,\n, \ldots$ are indices along $T^{1,1}$, while
hatted indices $\hat\mu,\hat\nu,\dots$ refer to cobasis directions on $M^d\times T^{1,1}$.}
\bea
\nonumber \cf_{ab} &=& F_{ab} \ \coloneqq \ \big( \diff A + A\wedge A
\big)_{ab} \qquad\textrm{for}\quad a,b\=1, \ldots, d \ ,\\[4pt]
\nonumber \cf_{a \m}  &=& \big(\diff X_{\m}\big)_{a}
+\com{A_{a}}{X_{\m}} \ \eqqcolon \ D_{a} X_{\m}
\qquad\textrm{for}\quad \m\=1, \ldots, 5 \ ,\\[4pt]
\nonumber \cf_{12} &=& \com{X_1}{X_2} -2 X_5+ \mbox{$\frac{3}{2}$} \im
I_6 \ , \ \ \ \ \cf_{13} \= \com{X_1}{X_3} \ , \ \ \ \ \ \ \
\cf_{14}\= \com{X_1}{X_4} \ ,\\[4pt]
\nonumber \cf_{15} &=& \com{X_1}{X_5}+\mbox{$\frac{3}{2}$}\, X_2 \ , \
\ \ \ \ \ \ \ \ \ \   \ \ \ \cf_{23}\= \com{X_2}{X_3} \ , \ \ \ \ \ \
\ \cf_{24} \= \com{X_2}{X_4} \ ,\\[4pt]
\nonumber \cf_{25} &\=& \com{X_2}{X_5}-\mbox{$\frac{3}{2}$}\, X_1 \ ,
\ \ \ \ \ \ \ \ \ \ \   \ \ \ \cf_{34} \= \com{X_3}{X_4}-2
X_5-\mbox{$\frac{3}{2}$} \im I_6 \ ,\\[4pt]
 \cf_{35} &\=& \com{X_3}{X_5}+\mbox{$\frac{3}{2}$}\, X_4 \ , \ \ \ \ \
 \ \ \ \ \ \   \ \ \ \cf_{45} \= \com{X_4}{X_5}-\mbox{$\frac{3}{2}$}\,
 X_3 \ ,
\label{eq:field_strength_components}
\eea
where we denote by $D_{a}$ the covariant derivatives (still assuming that the Higgs fields depend only on the coordinates of the manifold $M^d$).  
The resulting Yang-Mills Lagrangian is given by 
\bea
 \mathcal{L}_{\mathrm{YM}} \= -\mbox{$\frac{1}{4}$}\, \sqrt{\hat{g}}\
 \mathrm{tr} \ \cf_{\hat{\m}\hat{\n}}\, \cf^{\hat{\m}\hat{\n}}
       \=-\mbox{$\frac{1}{4}$}\, \sqrt{\hat{g}}\ \mathrm{tr} \
       g^{\hat{\sigma}\hat{\m}} \, g^{\hat{\rho}\hat{\n}}\, 
                      \cf_{\hat{\m}\hat{\n}}\,
                      \cf_{\hat{\sigma}\hat{\rho}} \ ,
\eea
where we denote $\hat{g}= \mathrm{det}(g_{M^d})\, \mathrm{det}( g)$ with
$g_{M^d}$ the metric on $M^d$.
As the metric $g$ on the homogeneous space $T^{1,1}$ is given in
terms of an orthonormal coframe, the Lagrangian simply reads
\small
\bea
\label{LYM}
\nonumber    \mathcal{L}_{\mathrm{YM}} &=& -\frac{1}{2}\,
\sqrt{\hat{g}}\ \mathrm{tr}\ \Big(\, \frac{1}{2}\, F_{ab}\,
  F^{ab}+\sum_{\m=1}^{5} \, \lb D_{a}  
                                               X_{\m}\rb \lb
                                               D^{a}X_{\m}\rb \\ \nonumber
                                               && +\, \big(
                                               \com{X_1}{X_3}\big)^2+
                                               \big(
                                               \com{X_1}{X_4}\big)^2
                                               + \big(
                                               \com{X_2}{X_3}\big)^2 + \big( \com{X_2}{X_4}\big)^2
                              \\ && +\, \big(
                              \com{X_1}{X_2}-2X_5+\mbox{$\frac{3}{2}$}
                              \im I_6\big)^2 
                                                 + \big(
                                                 \com{X_3}{X_4}-2X_5-\mbox{$\frac{3}{2}$}
                                                 \im I_6\big)^2 \\
\nonumber                                       && +\, \big(
  \com{X_1}{X_5}+\mbox{$\frac{3}{2}$}\, X_2\big)^2 + \big(
  \com{X_2}{X_5}-\mbox{$\frac{3}{2}$}\, X_1\big)^2 		 	
                                                 +\big(
                                                 \com{X_3}{X_5}+\mbox{$\frac{3}{2}$}\,
                                                 X_4\big)^2 
                                                 +\big(
                                                 \com{X_4}{X_5}-\mbox{$\frac{3}{2}$}\,
                                                 X_3\big)^2 \, \Big) \
                                                 .
\eea
\normalsize

The corresponding action functional is given by 
\beq\label{SYM}
 S_{\mathrm{YM}} \= \int_{M^d\times T^{1,1}} \, \diff^{d +5}x \
 \mathcal{L}_{\mathrm{YM}} \ .
\eeq
Since the Higgs fields do not depend on coordinates on $T^{1,1}$, the integral over the 
coset space simply yields its volume $\mathrm{Vol}\lb
T^{1,1}\rb=\frac{16 \pi^3}{27}$ in the chosen metric $g$. 
Hence dimensional reduction over $T^{1,1}$ of the Yang-Mills
Lagrangian on $M^d\times T^{1,1}$ becomes
\beq
\mathcal{L}_{\mathrm{r}}\= \frac{16 \pi^3}{27} \,
\mathcal{L}_{\mathrm{YM}} \qquad\textrm{and}\qquad S_{\rm r}\= \int_{M^d}\,
\diff^d  x\ \mathcal{L}_{\mathrm{r}} \ ,
\eeq
which describes a Yang-Mills-Higgs theory on $M^d$. 
Of course, the result we are interested in is the concrete form of the Higgs contributions induced by the internal manifold. 
We have seen that imposing the compatibility condition of equivariance has ruled out many contributions to the connection 
and has fixed the form of the Lagrangian of the gauge theory as in
(\ref{LYM}). After choosing the decomposition of the representation $(m_1,m_2)$ of
the structure group, the only freedom that remains is the concrete realization of the allowed endomorphisms, represented by the arrows in the quiver. 
The instanton equations will require further relations among them, which shall be studied in the next section.  

Often the restrictions of $G$-equivariance are so strong that the explicit evaluation of the Higgs contributions is significantly simplified. 
For this, we consider as an example the special solution of the
$T^{1,1}$ quiver constraints given
by the rectangular $A_{m_1}\oplus A_{m_2}$ quiver 
\beq
    \begin{tikzpicture}[->,scale=1.9]
    \node (a) at (0,0) {$\mbf{(0,0)}$};
    \node (b) at (1,0) {$\mbf{(1,0)}$};
    \node (c) at (2,0) {$\mbf{(2,0)}$};
    \node (d) at (3,0) { \ $\ldots$ \ };
    \node (e) at (4,0) {$\mbf{(m_1,0)}$};

    \node (g) at (0,1) {$\mbf{(0,1)}$};
    \node (h) at (1,1) {$\mbf{(1,1)}$};
    \node (i) at (2,1) {$\mbf{(2,1)}$};
    \node (j) at (3,1) { \ $\ldots$ \ };
    \node (k) at (4,1) {$\mbf{(m_1,1)}$};

    \node (m) at (0,2) {$\vdots$};
    \node (n) at (1,2) {$\vdots$};
    \node (o) at (2,2) {$\vdots$};
    \node (p) at (3,2) {};
    \node (q) at (4,2) {$\vdots$};

    \node (s) at (0,3) {$\mbf{(0,m_2)}$};
    \node (t) at (1,3) {$\mbf{(1,m_2)}$};
    \node (u) at (2,3) {$\mbf{(2,m_2)}$};
    \node (v) at (3,3) { \ $\ldots \ $};
    \node (w) at (4,3) {$\mbf{(m_1,m_2)}$};

    \path [thick](b) edge node {} (a);
    \path [thick](c) edge node {} (b);
    \path [thick](d) edge node {} (c);
    \path [thick](e) edge node {} (d);

    \path [thick](h) edge node {} (g);
    \path [thick](i) edge node {} (h);
    \path [thick](j) edge node {} (i);
    \path [thick](k) edge node {} (j);

    \path[dashed, thick, blue] (g) edge node {} (a);
    \path[dashed, thick, blue] (h) edge node {} (b);
    \path[dashed, thick, blue] (i) edge node {} (c);
    \path[dashed, thick, blue] (k) edge node {} (e);

    \path[dashed, thick, blue] (m) edge node {} (g);
    \path[dashed, thick, blue] (n) edge node {} (h);
    \path[dashed, thick, blue] (o) edge node {} (i);
    \path[dashed, thick, blue] (q) edge node {} (k);

    \path[dashed, thick, blue] (s) edge node {} (m);
    \path[dashed, thick, blue] (t) edge node {} (n);
    \path[dashed, thick, blue] (u) edge node {} (o);
    \path[dashed, thick, blue] (w) edge node {} (q);

    \path [thick](t) edge node {} (s);
    \path [thick](u) edge node {} (t);
    \path [thick](v) edge node {} (u);
    \path [thick](w) edge node {} (v);

    \draw [->, dotted, thick, red](a) to [out=120,in=150,looseness=7] (a);
    \draw [->, dotted, thick, red](b) to [out=120,in=150,looseness=7] (b);
    \draw [->, dotted, thick, red](c) to [out=120,in=150,looseness=7] (c);
    \draw [->, dotted, thick, red](e) to [out=120,in=150,looseness=7] (e);
    \draw [->, dotted, thick, red](g) to [out=120,in=150,looseness=7] (g);
    \draw [->, dotted, thick, red](h) to [out=120,in=150,looseness=7] (h);
    \draw [->, dotted, thick, red](i) to [out=120,in=150,looseness=7] (i);
    \draw [->, dotted, thick, red](k) to [out=120,in=150,looseness=7] (k);
    \draw [->, dotted, thick, red](s) to [out=120,in=150,looseness=7] (s);
    \draw [->, dotted, thick, red](t) to [out=120,in=150,looseness=7] (t);
    \draw [->, dotted, thick, red](u) to [out=120,in=150,looseness=7] (u);
    \draw [->, dotted, thick, red](w) to [out=120,in=150,looseness=7] (w);
    \end{tikzpicture}
\label{fig4}
\eeq
with one loop at each
vertex from the extra vertical component.
This quiver arises if (\ref{eq46}) is solved by imposing 
(\ref{eq:cp1cp1_cond}) and by demanding that the Higgs field $\phi^{(3)}$ is
diagonal. It is then possible to exploit a grading of the connection, similarly to that of~\cite{LPS06}, which greatly reduces the number of contributions. With the abbreviations
$\phi^{(1)}_{i+1\, \a}\coloneqq  \phi^{(1)}_{i\, \a,i+1\, \a}$,
$\phi^{(2)}_{i\, \a+1}\coloneqq  \phi^{(2)}_{i\, \a,i\, \a+1}$ and
$\phi^{(3)}_{i\a}\coloneqq \phi^{(3)}_{i\a,i\a}$, 
the non-vanishing block components of the connection read 
\bea
\label{eq58}
\nonumber \ca^{i\alpha, i\alpha} &=& A^{i\alpha} + \phi_{i
  \alpha}^{(3)}\, e^5 +c_{i\a}\, \id_{k_{i\alpha}}\, a
\qquad\textrm{with}\quad c_{i\a}\=m_1-m_2-2i+2\a \ ,\\[4pt]
\nonumber \ca^{i\,\alpha,i+1\,\alpha}&=& \phi^{(1)}_{i+1\, \alpha} \lb
e^1-\im e^2\rb \= -\lb \ca^{i+1\,\alpha,i\,\alpha}\rb^{\dagger} \ ,\\[4pt]
          \ca^{i\,\alpha,i\,\alpha +1}&=&\phi^{(2)}_{i\, \alpha +1} \lb
          e^3-\im e^4\rb \=-\lb \ca^{i\,\alpha +1, i\,\alpha}
          \rb^{\dagger} \ ,
\eea
where $A^{i\a}$ is a connection on the vector bundle $E_{i\a}\to M^d$
with curvature $F^{i\a}=\diff A^{i\a}+A^{i\a}\wedge A^{i\a}$. From the field strength
$\cf^{i\alpha,j\beta}= \diff \ca^{i\alpha,j\beta}+\sum_{l, \gamma}\, \ca^{i\alpha, l\gamma}\wedge\ca^{l\gamma,j\beta}$
we compute the reduced action functional and obtain\footnote{
Alternatively, we may substitute directly into (\ref{LYM}).} (see
Appendix~A for details)
\small
\bea \nonumber
 S_{\mathrm{r}} &=& \frac{16\pi^3}{27}\, \int_{M^d}\, \diff^d x \ \sqrt{\det(g_{M^d})} \
 \sum_{i=0}^{m_1} \ 
                      \sum_{\a =0}^{m_2}\, \mathrm{tr}
                      \left \{ \mbox{$\frac{1}{4}$} \, F_{ab}^{i\a}
                        \,^{\dagger} \, F^{i\a \, ab} + 
                      \mbox{$\frac{1}{2}$}\, \big(
                      D_{a}\phi^{(3)}_{i\a}\big)\, \big( D^{a}\phi^{(3)}_{i\a}\big)^{\dagger}\right.\\
\nonumber        &&\left.+\, \big(
  D_{a}\phi^{(1)}_{i\a}\big)^{\dagger} \, \big( D^{a}\phi^{(1)}_{i\a}\big)
                          + \big( D_{a}\phi^{(1)}_{i+1\, \a}\big)\,
                          \big( D^{a}\phi^{(1)}_{i+1\, \a}\big)^{\dagger}
                          + \big( D_{a}\phi^{(2)}_{i\a}\big)^{\dagger}
                          \, \big( D^{a}\phi^{(2)}_{i\a}\big) + \big(
                          D_{a}\phi^{(2)}_{i\, \a +1}\big)\, \big(
                          D^{a}\phi^{(2)}_{i\, \a +1}\big)^{\dagger}
                          \right.\\
\nonumber        &&\left.+\, 2 \big|\phi^{(1)}_{i\a}\,^{\dagger} \,
  \phi_{i\a}^{(1)} - \phi^{(1)}_{i+1 \, \a} \, \phi^{(1)}_{i+1 \, \a}{}^{\dagger}
                         + \im \phi^{(3)}_{i\a}+\mbox{$\frac{3}{4}$}\, c_{i\a}\, \id_{k_{i\a}}\big|^2 \right. \\
\nonumber       &&\left. +\, 2 \big| \phi^{(2)}_{i\a}\,^{\dagger} \, \phi_{i\a}^{(2)} - \phi^{(2)}_{i+1\, \a} \, \phi^{(2)}_{i+1\, \a}{}^{\dagger}
                         + \im \phi^{(3)}_{i\a}-\mbox{$\frac{3}{4}$}\,
                         c_{i\a} \, \id_{k_{i\a}}\big|^2 \right. \\
\nonumber       &&\left. +\, 2 \big|\phi^{(1)}_{i+1\, \a}\,
  \phi^{(2)}_{i+1\, \a +1}-\phi^{(2)}_{i\, \a +1} \, \phi^{(1)}_{i+1\, \a +1} \big|^2
                     + 2 \big|\, \big(\phi^{(1)}_{i\, \a -1} \, \phi^{(2)}_{i\a}-
                     \phi^{(2)}_{i-1\, \a}\, \phi^{(1)}_{i\a} \big)^\dag
                     \, \big|^2 \right.\\
\nonumber       && \left. +\, 2 \big| \phi^{(2)}_{i\a} \,^{\dagger} \,
  \phi^{(1)}_{i+1\, \a -1} - \phi^{(1)}_{i+1\, \a} \,
  \phi^{(2)}_{i+1\, \a}{}^{\dagger}\big|^2 + 2 \big| \, \big(
  \phi^{(2)}_{i-1\, \a
    +1}{}^{\dagger} \, \phi^{(1)}_{i\a}
                    -\phi^{(1)}_{i\, \a +1}\, \phi^{(2)}_{i\, \a
                      +1}{}^{\dagger}\big)^\dag\, \big|^2 \right. \\
\nonumber     &&\left.+\, \big|\, \big(\phi^{(1)}_{i\a}\,
  \phi^{(3)}_{i\a}-\phi^{(3)}_{i-1\, \a}\,
  \phi^{(1)}_{i\a}-\mbox{$\frac{3}{2}$} \im
  \phi^{(1)}_{i\a}\big)^\dag\, \big|^2
                +\big|\phi^{(1)}_{i+1\, \a}\,
                \phi^{(3)}_{i+1\, \a}-\phi^{(3)}_{i\a}\,
                \phi^{(1)}_{i+1\, \a}-\mbox{$\frac{3}{2}$} \im
                \phi^{(1)}_{i+1\, \a} \big|^2 \right.\\
\label{eq68}  && \left.+\, \big|\, \big( \phi^{(2)}_{i\a}\,
  \phi^{(3)}_{i\a}-\phi^{(3)}_{i\, \a -1}\, \phi^{(2)}_{i\a}
  -\mbox{$\frac{3}{2}$} \im \phi^{(2)}_{i\a} \big)^\dag\, \big|^2
                + \big|\phi^{(2)}_{i\, \a +1}\, \phi^{(3)}_{i\, \a
                  +1}-\phi^{(3)}_{i\a}\, \phi^{(2)}_{i\, \a
                  +1}-\mbox{$\frac{3}{2}$} \im \phi^{(2)}_{i\, \a
                  +1}\big|^2 \right \} \ ,
\eea
\normalsize
where we have
used the abbreviation $|\Phi|^2:=\Phi\, \Phi^\dag$ and the covariant derivatives from (\ref{eq:field_strength_components}) take the form 
\bea
D \phi^{(1)}_{i+1\, \a} &=&  \diff \phi^{(1)}_{i+1\, \a} + A^{i\a}\,
\phi^{(1)}_{i+1\, \a}-\phi^{(1)}_{i+1\, \a}\, A^{i+1\, \a} \ , \nonumber\\[4pt]
D \phi^{(2)}_{i\, \a +1} &=& \diff \phi^{(2)}_{i\, \a
  +1}+A^{i\a}\, \phi^{(2)}_{i\, \a +1}-\phi^{(2)}_{i\, \a +1}\, A^{i\, \a
  +1} \ , \nonumber\\[4pt]
D \phi^{(3)}_{i\a} &=& \diff \phi^{(3)}_{i\a}+A^{i\a}\,
\phi^{(3)}_{i\a}-\phi^{(3)}_{i\a}\, A^{i\a} \ . 
\eea
The most prominent difference between 
this reduced action functional and that of \cite{LPS06} is, of course, the
appearance of the third Higgs field $\phi^{(3)}$ due to
the remaining vertical component. As mentioned before, the radii of
the two spheres, which occur as moduli in the action of \cite{LPS06},
have been fixed here to numerical values by the Sasaki-Einstein condition. From (\ref{eq68}) one recovers indeed 
the action functional corresponding to dimensional reduction over the K\"ahler
coset space\footnote{
For the correctly fixed values of the radii, see Appendix~A.} $\C P^1 \times \C P^1$ 
by taking the limit $\phi^{(3)}= -\frac{3\im}{4}\,
\big(
\Upsilon^{(1)}+\Upsilon^{(2)}\big)$. Note that
naively setting the additional Higgs field to zero leads to a
different action functional. 

The next task is to study the equations of motion, and in particular
determine the vacua that are described by the Lagrangian
(\ref{LYM}). For this, we have to 
solve the Yang-Mills equations on $M^d\times T^{1,1}$, which is
simplified by the existence of Killing spinors on the Sasaki-Einstein
manifold $T^{1,1}$ because solutions of the instanton 
equation (\ref{insta:eq}) also satisfy the Yang-Mills equations on
$T^{1,1}$ in this
instance~\cite{noelle12}. Furthermore, it is even 
more convenient to work in even dimensions over the corresponding
Calabi-Yau metric cone $C(T^{1,1})$, as
one can then solve the {Hermitian Yang-Mills equations}, 
which imply the instanton equations. This is the subject of the remainder of this paper. 


\section{Instantons on the conifold}
\subsection{Geometry of the cone $C(T^{1,1})$}

As a metric cone over a Sasaki-Einstein manifold, the conifold $C( T^{1,1})$ is by contruction a (non-compact) Calabi-Yau manifold, 
so that its Riemannian holonomy is contained in $\mathrm{SU}(3)$ and it is Ricci-flat. In contrast to the common framework of 
compactifications on orbifolds in string theory, 
the conifold cannot be described as a global quotient $\C^3/\Gamma$ by
a discrete subgroup $\Gamma\subset\mathrm{SU}(3)$ because it is not
flat. On the other hand, it also admits a description as a toric variety
(see e.g.~\cite{Davies11}) which is described in Cox homogeneous
coordinates as the quotient space\footnote{
In this description, naive dimensional reduction of Yang-Mills theory
over the conifold yields a quiver gauge theory based on the
Klebanov-Witten quiver~\cite{T11}.
}
\beq
C\big(T^{1,1} \big) \ \simeq \ \big(\C^4\setminus\mathcal{Z}\big) \, \big/ \, \C^*
\eeq
with weights $(1,1,-1,-1)$, i.e. one identifies
$(z_1,z_2,z_3,z_4)\sim (\lambda\, z_1, \lambda \, z_2, \lambda^{-1}\,
z_3, \lambda^{-1}\, z_4)$ for $\lambda \in \C^*$, where $\mathcal{Z}$ is
the union of the loci of points $(z_1,z_2,0,0)\neq(0,0,0,0)$ and $(0,0,z_3,z_4) \neq(0,0,0,0)$. 
Therefore one cannot study translationally invariant instantons in this case, which would lead to a quiver gauge theory generated by equivariance conditions with 
respect to the discrete group $\Gamma$, see e.g.~\cite{Orbifolds}. Hence we shall briefly consider the geometry of 
the metric cone before we proceed to the description of instanton solutions on it.

By definition, the metric of the conifold is the warped product
\bea
\label{eq69}
g_{\rm con} \= r^2 \, g + \diff
r \otimes \diff r \= r^2 \, \sum_{\m=1}^6\, e^{\m} \otimes e^{\m}
 \= \e^{2\tau}\, \sum_{\m=1}^6 \, e^{\m} \otimes e^{\m}
\eea
with radial coordinate $r\in\R_{>0}$, where this equation establishes a conformal equivalence between the
cone metric and the metric on the cylinder by setting
\beq
e^{6}\= \mbox{$\frac{1}{r}$} \, \diff r \= \diff \tau \qquad\textrm{with}\quad \tau \
\coloneqq \ \log(r) \ . 
\eeq
The K\"ahler form $\Omega ( \cdot, \cdot)=g_{\rm con}( J\cdot, \cdot)$
(using the cylinder metric\footnote{
In what 
follows the descriptions with cone metric and with cylinder metric are considered equivalent.}) is given by
\beq
\label{eq70}
\Omega \= r^2\, \omega^3 + r^2\, e^5\wedge e^6 \= r^2 \lb
e^{12}+e^{34}+e^{56}\rb \ ,
\eeq
which is closed due to the defining Sasaki-Einstein relations (see
Appendix~A for details), and the holomorphic 1-forms are 
\beq
\Theta^a \ \coloneqq \ e^{2a-1}+\im e^{2a} \qquad\textrm{with}\quad J
\Theta^a \= \im \Theta^a \qquad\textrm{for}\quad a\=1,2,3 \ .
\eeq

By rescaling the forms 
\beq
\et^{\m} \ \coloneqq \ r\, e^{\m} \ , \qquad
\diff \et^{\m}\= r \, \diff e^{\m} - e^{\m} \wedge \diff r \ , 
\eeq
one obtains an orthonormal cobasis and structure equations with respect to the cone metric. 
The connection matrix for the Levi-Civita connection is given by
\beq
\label{gamma_cone}
\diff \begin{pmatrix}
       \et^1- \im \et^2\\
       \et^3- \im \et^4\\
       \et^5 -\im \et^6
      \end{pmatrix}
\= -\begin{pmatrix}
     2 a -\frac{1}{2}\im e^5   & 0    & \im \lb e^1-\im e^2\rb \\
     0                    & -2a -\frac{1}{2} \im e^5  & \im \lb e^3-\im e^4\rb\\
     \im \lb e^1 +\im e^2\rb & \im \lb e^3+ \im e^4\rb & \im e^5       
   \end{pmatrix}
\wedge
\begin{pmatrix}
       \et^1- \im \et^2\\
       \et^3- \im \et^4\\
       \et^5-\im \et^6
      \end{pmatrix} \ .
\eeq
Being a Calabi-Yau threefold, the conifold has holonomy group $\mathrm{SU}(3)$, and a 
calculation of the curvature of the Levi-Civita connection 
(see Appendix~A) shows that it is valued in the Lie subalgebra $\mathfrak{su}(2) \subset \mathfrak{su}(3)$ 
and solves the instanton equation~(\ref{insta:eq}).\footnote{
The existence of an $\su$-structure on a (real) six-dimensional Calabi-Yau manifold implies that it has
an almost product structure, see e.g.~\cite{su2_structure}. The conifold locally looks like a product of $\mathbb{R}_{>0} \times S^1$ with 
$\C P^1 \times \C P^1$.} On the other hand, declaring again all 
terms apart from those containing the form $a$ as torsion, one obtains 
a $\mathrm{U}(1)$ connection which is simply the lift of the canonical
connection on $T^{1,1}$ to the cone; it is clearly still an instanton.
Consequently, we have now two instanton solutions to start from in our
construction.

For this, let us adapt the approach used 
on $T^{1,1}$ more generally to the metric cone $C(T^{1,1})$. Given an
instanton $\Gamma=\Gamma^i \, I_i$ with generators
$I_i\in\mathfrak{su}(2)\subset \mathfrak{u}(k)$ for $i=6,7,8$, the ansatz for the connection reads
\beq
\ca\= \Gamma +X_{\m} \, e^{\m} \ .
\eeq
From the structure equations $\diff e^{\m}=-\Gamma^{\m}_{\ \n}\wedge
e^{\n}+\frac{1}{2}\, T_{\rho \sigma}^{\m}\, e^{\rho \sigma}$ and the Maurer-Cartan equation
$\diff e^{\m}=-\frac{1}{2}\, f_{\rho \sigma}^{\m}\, e^{\rho
  \sigma}$, it follows that $\lb \Gamma^{\m}_{\ \n}\rb_i=f^{\m}_{i\n}$ and the curvature yields
\bea
\cf \= \cf_{\Gamma} + \Gamma^i \lb \com{I_i}{X_{\m}}-f_{i \m}^{\n}\,
X_{\n}\rb \wedge e^{\m} +\mbox{$\frac{1}{2}$} \lb
\com{X_{\m}}{X_{\n}}+T^{\sigma}_{\m \n}\, X_{\sigma}\rb e^{\m\n}
            +\diff X_{\m} \wedge e^{\m} \ ,
\label{eq:comp_conn}            
\eea
where $\cf_{\Gamma}\coloneqq \diff\Gamma +\Gamma \wedge \Gamma$. Now choose the endomorphisms $X_{\m}$ such that the differentials
$\diff X_{\m}$ do not yield contributions containing the forms
$\Gamma^i$; in particular, this holds for the case of constant
matrices and for spherically symmetric matrices $X_{\m}=X_{\m}( r)$ as
instanton solutions that we consider below. Then the equivariance condition is exactly expressed by the 
vanishing of the second term, 
\beq
\com{I_i}{X_{\m}} \= f_{i \m}^{\n}\, X_{\n} \ ,
\eeq
which generates the quiver.

Given the compatible connection and its curvature (\ref{eq:comp_conn}), one can obtain instanton solutions by using 
the K\"ahler form $\Omega$ for the formulation of the \emph{Hermitian Yang-Mills equations}~\cite{Popov09}
\bea
\cf^{2,0}\= 0 \= \cf^{0,2} \ , \ \ \ \ \ \ \ \Omega \haken \cf^{1,1} \=
0 \ ,
\label{eq:HYM}
\eea
where $\cf=\cf^{2,0}+\cf^{1,1}+\cf^{0,2}$ refers to the decomposition
into holomorphic and antiholomorphic parts with respect to the complex
\mbox{structure $J$}, so that the 
first equation means that the field strength is invariant under the action of $J$. These
equations can be regarded as stability conditions\footnote{
This pertains to compact Calabi-Yau manifolds;
on the conifold they are simply a set of additional real differential
equations imposed on a Hermitian connection.} on holomorphic
vector bundles and are sometimes referred to as \emph{Donaldson-Uhlenbeck-Yau equations}~\cite{Donaldson85,UYau}; they imply the instanton equation~(\ref{insta:eq}).


\subsection{Yang-Mills flows}
When applying this construction to the lifted canonical connection
$\Gamma = I_6 \, a$ on $C(T^{1,1})$, the only difference from before
that one has to take into account is the
 additional radial coordinate giving rise to one further endomorphism $X_6$,
\bea
\label{eq71}
\ca \= I_6 \, a+ \sum_{\m=1}^5\, X_{\m} \, e^{\m} +X_6\, e^6 \ . 
\eea
We are interested in spherically symmetric instanton solutions, i.e.
those endomorphisms $X_{\m}=X_{\m}(\tau)$ which depend only on the radial coordinate
$r=\e^\tau$. 
After implementing the equivariance conditions\footnote{
Here we write $\phi^{(a)}\coloneqq \frac{1}{2}\lb X_{2a-1}+\im X_{2a}\rb$ for $a=1,2,3$.}
\bea
\com{I_6}{\phi^{(1)}}\=2 \phi^{(1)} \ , \qquad \com{I_6 }{
  \phi^{(2)}}\=-2\phi^{(2)} \ , \qquad \com{I_6}{ \phi^{(3)}}\=0 \ ,
\eea
which are identical to those over $T^{1,1}$, the field strength is given by
\bea
\label{eq72}
\cf \= \cf_{T^{1,1}} + \sum_{\m=1}^5\, \Big(\, \com{X_{\m}}{X_6}-
\frac{\diff X_{\m}}{\diff\tau} \, \Big)\, e^{\m6} \ ,
\eea
where $\cf_{T^{1,1}}$ denotes the curvature we have already derived over $T^{1,1}$ with components $\cf_{\m\n}$ 
from (\ref{eq:field_strength_components}).

Evaluating the holomorphicity condition of the Hermitian Yang-Mills equations (\ref{eq:HYM}) 
in terms of the holomorphic 1-forms $\Theta^a$ leads to four first
order ordinary differential equations
\bea
\nonumber \frac{\diff X_1}{\diff\tau} &=&   -\frac{3}{2} \,X_1 +\com{X_1}{X_6}+\com{X_2}{X_5} \ , \ \ \ \ 
          \frac{\diff{X}_2}{\diff\tau} \=  -\frac{3}{2} \,X_2 +\com{X_2}{X_6}-\com{X_1}{X_5} \ ,\\[4pt]
          \frac{\diff{X}_3}{\diff\tau}  &=&  -\frac{3}{2} \,X_3 +\com{X_3}{X_6}+\com{X_4}{X_5} \ , \ \ \ \ 
          \frac{\diff{X}_4}{\diff\tau} \= -\frac{3}{2}\,X_4 +\com{X_4}{X_6}-\com{X_3}{X_5} \ ,
\label{eq77}
\eea
together with the constraints
\beq
\com{X_1}{X_3} \= \com{X_2}{X_4} \qquad\textrm{and}\qquad \com{X_1}{X_4} \= -\com{X_2}{X_3} \ .
\label{eq78}
\eeq
The remaining stability condition $\Omega \haken \cf =0$ yields the flow equation for $X_5$ given by
\beq
\frac{\diff{X}_5}{\diff\tau} \=  -  4 X_5 +\com{X_1}{X_2}+ \com{X_3}{X_4} +\com{X_5}{X_6} \ .
\label{eq:flow5}
\eeq
Inserting these equations into the action functional (\ref{SYM}) over
the cylinder $\R_{>0}\times T^{1,1}$ leads to cancellations of many
contributions involving the Higgs potential, as to be expected from a vacuum solution on the Higgs branch of the quiver gauge theory.
Moreover, one can see that the conditions imposed by the Hermitian Yang-Mills equations induce relations on the quiver:  
The contraints (\ref{eq78}) are cast into the quiver relation
\beq
\com{\phi^{(1)}}{ \phi^{(2)}}\=0 \ ,
\label{eq:comm}
\eeq
i.e. the commutativity of the Higgs fields $\phi^{(1)}$ and $\phi^{(2)}$ follows naturally as a consequence of the Hermitian Yang-Mills equations. 
In the simplified example of the $A_{m_1}\oplus A_{m_2}$ quiver with
vertex loops, which admits a grading of the connection, this implies
commutativity of the quiver arrows around 
the rectangular lattice, $\phi^{(1)}_{i+1\,\a}\, \phi^{(2)}_{i+1\,\a +1}= \phi^{(2)}_{i\, \a +1}\, \phi^{(1)}_{i+1\, \a +1}$. One can also directly observe the vanishing 
of the corresponding contributions to the Higgs potential in the action functional (\ref{eq68}). 

Before we describe the general solutions to these flow equations under the given constraints, we consider the case of constant 
endomorphisms. When the matrices $X_{\m}$ do not depend on $r$, the radial coordinate enters the framework just as a parameter that labels 
different copies of $T^{1,1}$ (as a foliation of the six-dimensional cone into copies of the underlying Sasaki-Einstein manifold along 
the preferred direction $r$). Therefore the examination of constant endomorphisms corresponds to studying instanton solutions for 
the original five-dimensional situation.\footnote{
The matrix $X_6$ can always be set to zero via a real gauge
transformation, see e.g.~\cite{S5}.}
 The flow equations turn exactly into the conditions (\ref{eq_correspondence})
 which arose as additional equivariance relations in the limit where
 the total space of the Sasakian fibration $T^{1,1}$ degenerates to its base $\C P^1 \times \C P^1$; they lead 
to the vanishing of most terms in the Lagrangian (\ref{LYM}) of the
quiver gauge theory on $T^{1,1}$. As discussed before, a solution to
these equations is given by
the choice $X_5 = -\frac{3\im}{4}\, \big(
\Upsilon^{(1)}+\Upsilon^{(2)}\big)$, which
shows that the quiver gauge theory on $\C P^1 \times \C P^1$ from~\cite{LPS06} for the 
appropriate values of the radii $R_l$ is not only contained in our
description but even automatically realizes a solution of the Hermitian Yang-Mills equations on the conifold.

\subsection{Instanton moduli spaces}

In solving the generic case of spherically symmetric instantons given by solutions to the flow equations (\ref{eq77}) and (\ref{eq:flow5}) under the derived constraints, one encounters 
 Nahm-type equations describing the radial dependence of the matrices $X_{\m}$. Hence one can apply techniques similar to those that have been used 
for the description of the hyper-K\"ahler structure of the moduli space of the (original) Nahm equations, see
in particular~\cite{Kronheimer90,Donaldson84}. In \cite{MS15}, instantons arising from the 
Hermitian Yang-Mills equations on Calabi-Yau cones of any dimension have been studied using these methods, and it was shown that the equations 
which describe the moduli space do not depend on the concrete Sasaki-Einstein manifold under consideration but only on its dimension; 
thus our equations for the moduli space of Hermitian Yang-Mills instantons on the conifold can be included in that treatment.

The flow equations can be brought to a form similar to the Nahm equations by setting 
\beq
X_i\= \e^{-\frac{3}{2}\, \tau}\ W_i \ , \quad i\=1,2,3,4 \qquad\textrm{and}\qquad
X_j\=\e^{-4\tau}\ W_j \ , \quad j\=5,6
\eeq
in order to eliminate the linear terms. Changing again the coordinate to 
\beq
s\= \mbox{$\frac{1}{4}$}\, \e^{-4\tau}\= \mbox{$\frac{1}{4}$}\, r^{-4} \ \in \ \R_{>0}
\eeq
and writing 
\bea
Z_1 \ \coloneqq \ \mbox{$\frac{1}{2}$} \lb W_{1}+\im W_{2}\rb \ , \ \ \ \ Z_2 \ \coloneqq \ \mbox{$\frac{1}{2}$} \lb W_3+\im W_4\rb \ , \ \ \ \  Z_3 \ \coloneqq \ \mbox{$\frac{\im}{2}$} \lb  W_5+\im W_6\rb \ ,
\label{eq81}
\eea
we arrive at the set of equations
\bea
\label{eq:cplx} \frac{\diff Z_1}{\diff s} &=& 2\com{Z_1}{Z_3} \ , \ \ \ \ \ \ \ \frac{\diff Z_2}{\diff s} \= 2\com{Z_2}{Z_3} \ ,\\[4pt]
          \frac{\diff Z_3}{\diff s}+\frac{\diff Z_3^{\dagger}}{\diff
            s} &=& 2
          (-s)^{-{5}/{4}} \, \big( \com{Z_1}{Z_1^{\dagger}}+\com{Z_2}{
            Z_2^{\dagger}} \big) - 2\com{Z_3}{Z_3^{\dagger}} 
\label{eq:re}
\eea
together with the constraints
\bea 
\com{I_6}{Z_1}\= 2Z_1 \ , \ \ \ \ \ \ \com{I_6}{Z_2}\=-2 Z_2 \ , \ \ \ \ \
\ \com{I_6}{Z_3}\=0 \ , \ \ \ \ \ \ \com{Z_1}{Z_2} \= 0 \ .
\label{eq83}
\eea
We therefore turn our attention to the moduli space of solutions to the equations~(\ref{eq:cplx})--(\ref{eq:re}) subject to the equivariance constraints and quiver relations from \eqref{eq83}; we refer to the two equations (\ref{eq:cplx}) as the \emph{complex equations} and to the equation (\ref{eq:re}) as the
\emph{real equation}. The complex equations and the constraints are
invariant under the complex gauge transformations~\cite{Donaldson84}
\bea
\nonumber Z_{a}&\longmapsto &g\cdot Z_a\= g\, Z_{a}\, g^{-1} \ , \ \ \ a\=1,2 \ , \\[4pt]
Z_3 &\longmapsto & g\cdot Z_3 \=g\, Z_3\, g^{-1} +\frac{1}{2} \,
\frac{\diff g}{\diff s} \, g^{-1}
\label{eq84}
\eea 
for arbitrary smooth functions $g: \R_{>0}\to \Gcal_\C \subset
\mathrm{GL}(k,\C)$, where $\Gcal$ is the subgroup of ${\rm U}(k)$
which stabilizes the generator $I_6$ under the adjoint action. This observation motivates a description of the moduli space of the 
flow equations in terms of a K\"ahler quotient construction, involving
an infinite-dimensional space of connections and an
infinite-dimensional gauge group, or equivalently in terms of
adjoint orbits. For instantons on Calabi-Yau cones the two approaches
have been carried out in~\cite{MS15}, whose results we will adapt to
our setting together with the subsequent implementations of the equivariance
constraints considered in~\cite{Lechtenfeld:2014fza,S5}.

For this, let $\mathbb{A}^{1,1}$ be the space of endomorphisms $Z_a$
satisfying the complex equations (\ref{eq:cplx}) and the constraints
(\ref{eq83}); the complexified gauge group $\widehat{\mathcal{G}}_\C$ acts on
$\mathbb{A}^{1,1}$ according to (\ref{eq84}). This space is naturally
an infinite-dimensional K\"ahler manifold with a gauge-invariant
metric and symplectic form~\cite{MS15,S5}. The corresponding moment
map $\mu:\mathbb{A}^{1,1}\to \widehat{\mathfrak{g}}$ is defined by
\bea
\mu\big(Z,Z^\dag\big)\= \frac{\diff Z_3}{\diff s}+\frac{\diff Z_3^{\dagger}}{\diff
            s} - 2
          (-s)^{-{5}/{4}} \, \big( \com{Z_1}{Z_1^{\dagger}}+\com{Z_2}{
            Z_2^{\dagger}} \big) + 2\com{Z_3}{Z_3^{\dagger}} \ ,
\eea
where $\widehat{\mathfrak{g}}$ is the Lie algebra of
$\widehat{\mathcal{G}}$ consisting of infinitesimal (real) gauge
transformations which commute with the generator $I_6$. The moment map
thus connects the space of solutions to the complex equations with the remaining real
equation, so that the moduli space $\man$ of Hermitian Yang-Mills instantons is
obtained by taking the K\"ahler quotient
\bea
\man \= \mu^{-1}(0) \, \big/ \, \widehat{\mathcal{G}} \ .
\eea
This quotient can be related~\cite{MS15,S5} to the action of the
complexified gauge group on the set of stable points
$\mathbb{A}_{\mathrm{st}}^{1,1}\subset \mathbb{A}^{1,1}$ whose
$\ggc$-orbits intersect the zeroes of the moment map, so that the
moduli space is realised as the GIT quotient
\beq
\man \ \simeq\  \mathbb{A}_{\mathrm{st}}^{1,1} \, \big/ \, \ggc \ .
\eeq

Following again the treatments of the Nahm equations from~\cite{Kronheimer90,Donaldson84,Biquard96}
 and their extensions to our setting of six-dimensional conical
 instantons from~\cite{MS15,S5}, we rewrite
 the solutions of the flow equations by applying a complex gauge
 transformation (\ref{eq84}) which locally trivializes the matrix $Z_3$ as
\beq
 Z_3 \= -\frac{1}{2} \, g^{-1}\, \frac{\diff g}{\diff s} \ . 
\label{eq:trafo1}
\eeq
In this gauge, the complex equations (\ref{eq:cplx}) are solved by gauge transformations of
constant matrices $U_1$ and $U_2$ as
\beq
{Z}_1 \= g^{-1}\, U_1\, g \ , \ \ \ \ \ \  \ \ \ {Z}_2 \= g^{-1}\, U_2
\, g \ .
\label{eq:trafo2}
\eeq
In order to fulfill the constraints \eqref{eq83}, the matrices $U_1$
and $U_2$ must be mutually commuting and satisfy the equivariance
conditions $\com{I_6}{U_1}=2U_1$ and $\com{I_6}{U_2}=-2U_2$. By generalizing Donaldson's treatment
of the ordinary Nahm equations~\cite{Donaldson84}, one can show that
these gauge fixed solutions fulfill the remaining real equation
\eqref{eq:re}, i.e. there exists a unique path $g(s)$ for
$s\in\R_{>0}$ which satisfies the real equation~\cite{MS15}.

Finally, we need to impose suitable boundary conditions. One
can adapt Kronheimer's asymptotics~\cite{Kronheimer90}  for
the solutions to the flow equations on the six-dimensional
cone~\cite{MS15,S5} as
\bea
\lim_{s\to\infty}\, W_{\m}(s) \= g_0\, T_{\m} \, g_0^{-1} 
\qquad\textrm{for}\quad \m\=1,\ldots,5 \ ,
\label{eq:asymptbc}\eea
where $g_0\in\Gcal$ and we have gauged away the scalar field $X_6$. Defining $V_a:=\frac12\, (T_{2a-1}+\im T_{2a})$ for
$a=1,2$, the constant boundary matrices satisfy
\bea
\com{I_6}{V_1}\= 2V_1 \ , \ \ \ \ \ \ \com{I_6}{V_2}\=-2 V_2 \ , \ \ \ \ \
\ \com{I_6}{T_5}\=0 \ , \ \ \ \ \ \ \com{V_1}{V_2} \= 0 \ .
\eea
The asymptotic boundary conditions \eqref{eq:asymptbc}
determine the singular behaviour of the
instanton connections $X_\mu$ as one approaches the conical
singularity at $r=0$ ($\tau\to-\infty$). On the other hand, one can
choose boundary conditions such that $X_\mu(\tau)\to0$ as
$\tau\to+\infty$, giving instantons that are \emph{framed} at infinity
in $\R_{\geq0}$, which implies that $W_\mu(s)$ has a limit as $s\to0$
whose value is completely determined by the solution of the first
order flow
equations with the boundary conditions \eqref{eq:asymptbc}. 

Following \cite{MS15,S5}, such solutions identify the moduli space $\mathcal{M}$ of the Hermitian Yang-Mills equations
on the metric cone in terms of adjoint orbits of the initial data
$T_{\m}$. This follows from \eqref{eq:cplx} which shows that the
solutions $Z_a(s)$ for $a=1,2$ each lie respectively in the same
adjoint orbit under the action of the complex Lie algebra 
$\mathfrak{gl}(k,\C)$ for all $s\in\R_{\geq0}$; by \eqref{eq:asymptbc}
they are
contained in the closures of the adjoint $\Gcal_\C$-orbits
${\mon_{V_a}}$ of $V_a$. By the above construction of local
solutions to the flow equations, for regular orbits ${\mon_{V_a}}$ the
map $Z_a(s)\mapsto Z_a(0)$ establishes a bijection
\bea
\man \ \simeq \ \mon_{V_1}\times \mon_{V_2}
\eea
which preserves the holomorphic symplectic structures. However, as
discussed in~\cite{Lechtenfeld:2014fza,S5}, the orbits ${\mon_{V_a}}$
are generally not regular and their closures generically coincide with
nilpotent cones consisting of nilpotent Lie algebra elements; that such singular
loci of fields arise is evident from the
solutions we found to the equivariance constraints in terms of graded
connections, for which the Higgs fields $\phi^{(1)}$ and $\phi^{(2)}$
are given by nilpotent matrices in
$\mathfrak{gl}(k,\C)$,
i.e. $\big(\phi^{(1)}\big)^{m_1+1}=0
=\big(\phi^{(2)}\big)^{m_2+1}$~\cite{LPS06}.

The moduli space $\man$ also parameterizes certain BPS configurations
of D-branes wrapping the conifold in Type~IIA string theory. For this,
we recall that the Hermitian Yang-Mills equations \eqref{eq:HYM} arise
as BPS equations for the (topologically twisted) maximally
supersymmetric Yang-Mills theory in six dimensions, which is obtained
by (naive) dimensional reduction of ten-dimensional $\mathcal{N}=1$
supersymmetric Yang-Mills theory to $C(T^{1,1})$, see
e.g.~\cite{Cirafici:2008sn}. In this way the equations \eqref{eq:HYM}
describe BPS bound states of D0--D2--D6 branes on the conifold, and
the moduli space $\man$ parameterizes spherically symmetric and equivariant
configurations thereof. In this context, the singularities of the moduli space $\man$ of ${\rm
  Spin}(4)$-equivariant instantons
corresponding to non-regular nilpotent orbits is reminiscent of those
of the moduli spaces of Hermitian Yang-Mills instantons on the (resolved) conifold which are
equivariant with respect to the maximal torus of the ${\rm SU}(3)$
holonomy group, see e.g.~\cite{Cirafici:2008sn,Cirafici:2012qc}.

On a more speculative front, we recall that moduli spaces of solutions
to the ordinary Nahm equations with Kronheimer's
boundary conditions also appear as Higgs moduli spaces of supersymmetric
vacua in $\Ncal=4$ supersymmetric Yang-Mills theory on the half-space
$\R^{1,2}\times\R_{\geq0}$ with generalized Dirichlet boundary
conditions~\cite{Gaiotto:2008sa}; these boundary conditions are realized by brane
configurations in which D$3$-branes transversally intersect D$5$-branes at the
boundary of $\R_{\geq0}$, which is the simple pole at $s=0$ of the
solutions to the Nahm equations. The flow equations in this case govern the
evolution of the Higgs fields of the $\Ncal=4$ gauge theory along the
direction $s\in\R_{\geq0}$, which represent the transverse
fluctuations of the D3-branes. It would be interesting to determine
whether the generalized Nahm equations \eqref{eq:cplx}--\eqref{eq:re}
can be derived analogously in terms of intersecting (pairs of)
D3-branes and D5-branes, with corresponding supersymmetric boundary
conditions in the worldvolume gauge theory, and hence if the instanton moduli
space $\man$ also parameterizes half-BPS states of certain D-brane
configurations in type~II string theory. 


\section{Conclusions}  

In this paper we examined dimensional reduction of
$\mathrm{Spin}(4)$-equivariant gauge theory over the coset space $T^{1,1}$ and characterized the compatible gauge connections in terms 
of representations of certain quivers. Special emphasis was placed on
a comparison with the quiver gauge theory obtained from dimensional
reduction over the K\"ahler coset space $\C P^1 \times \C P^1$
\cite{LPS06}, whose quiver representations are included
as special solutions in the more general framework over $T^{1,1}$. We showed that the Higgs fields depend on only one combined quantum number $c_{i\a}$
rather than on two individual monopole charges separately. In the
corresponding quivers we find more general arrows than the expected vertex loop modifications
of the rectangular $A_{m_1}\oplus A_{m_2}$ quiver. In addition, this
feature suggests an interpretation of the quiver gauge theory as that
of the double of an $A_{m_1+m_2}$ quiver with suitable combinations of
Higgs fields including multiplicities. The generic occurence of
doubles of
quivers resembles the situation which occurs in dimensional reduction
over quasi-K\"ahler coset spaces~\cite{Popov:2010rf}.

We studied the dimensional reduction of Yang-Mills theory and also
compared it to that associated to $\C P^1 \times \C P^1$
\cite{LPS06}. To study the Higgs branch of vacua of the quiver gauge theory, we made use 
of the special geometric structure of Sasaki-Einstein manifolds and
formulated a generalized instanton equation on the metric cone
$C(T^{1,1})$ by considering the Hermitian 
Yang-Mills equations. It was shown that the quiver gauge theory on the K\"ahler manifold $\C P^1 \times \C P^1$ (for the correct fixed values of the radii) is contained as 
an instanton solution
in the more general $T^{1,1}$ framework. The description of the moduli
space of Hermitian Yang-Mills instantons led to Nahm-type equations,
which we treated in terms of K\"ahler quotients and (nilpotent)
adjoint orbits, and argued to have a natural interpretation in terms of
BPS states of D-branes on the conifold.

\section*{Acknowledgements}
J.C.G.~thanks Marcus Sperling for helpful discussions. This work was partially supported by 
the Research Training Group RTG 1463, by the Grant LE 838/13 from the Deutsche Forschungs\-gemeinschaft (DFG, Germany), 
by the Consolidated Grant ST/L000334/1 from the UK Science and Technology Facilities Council (STFC), 
and by the Action MP1405 QSPACE from the European Cooperation in Science and Technology (COST).
\appendix
\section{Connections and curvatures}
\subsection{Connections on $T^{1,1}$}
The structure equations (\ref{eq13}) of the coset space $T^{1,1}$, with parameters fixed by the
 Sasaki-Einstein condition in \eqref{eq17}, can be expressed as
\beq
\diff \begin{pmatrix}
       e^1\\
       e^2\\
       e^3\\
       e^4\\
       e^5
      \end{pmatrix}
\= \begin{pmatrix}
    0              &2 \im a+\frac{1}{2}\, e^5  &      0 &   0 & -e^2\\
    -2 \im a -\frac{1}{2}\, e^5    & 0         &0  &0 & e^1\\
    0                      &     0 & 0 & -2 \im a +\frac{1}{2}\, e^5 & -e^4\\
   0              & 0 & 2 \im a -\frac{1}{2}\, e^5& 0 & e^3\\
  e^2 & -e^1 & e^3 & -e^4 &0
   \end{pmatrix}
\wedge 
\begin{pmatrix}
 e^1\\
 e^2\\
 e^3\\
 e^4\\
 e^5
\end{pmatrix} \ ,
\eeq   
from which we obtain the connection \mbox{1-forms} given in
(\ref{conn_lc}). The curvature tensor $R^{\m}_{\ \n}=\diff \Gamma^{\m}_{\ \n} + \Gamma^{\m}_{\ \sigma} \wedge \Gamma^{\sigma}_{\ \n}$ has the non-vanishing contributions
\bea
\nonumber R^{1}_{\ 2}&=& 3 e^{12}-2 e^{34} \ , \ \ \ R^{1}_{\ 3}\=
-e^{24} \ , \ \ \ R^{1}_{\ 4}\= e^{23} \ , \ \ \ R^{1}_{\ 5}\=e^{15} \
, \ \ \
R^{2}_{\ 3}\= e^{14} \ , \\[4pt]
         R^{2}_{\ 4}&=& -e^{13} \ , \ \ \ R^{2}_{\ 5}\= e^{25} \ , \ \
         \ R^{3}_{\ 4}\= -2e^{12}+3 e^{34} \ , \ \ \ R^{3}_{\
           5}\=e^{35} \ , \ \ \ R^{4}_{\ 5}\=e^{45} \ ,
\eea
 and hence
$\mathfrak{so}(5)$ holonomy. Expressing the curvature in components
$R^{\m}_{\ \n\l\kappa}$ and contracting to $R_{\l\kappa}=R^{\m}_{\ \l\m\kappa}$ 
yields the Ricci tensor (\ref{ricci_t11}).

The structure equations for the holomorphic forms $\Theta^1= e^1+\im e^2$ and $\Theta^2=e^3+\im e^4$ are
\bea
\nonumber \diff \Theta^1 &=& -2 \Theta^1 \wedge \theta^6 +2 \Theta^1 \wedge
\theta^5 \ , \ \ \ \ \ \ \ \
         \diff \bar{\Theta}{}^1 \= 2 \bar{\Theta}{}^1 \wedge \theta^6 - 2
         \bar{\Theta}{}^1 \wedge \theta^5
         \ , \\[4pt]
\nonumber \diff \Theta^2 &=& 2 \Theta^2 \wedge \theta^6 + 2 \Theta^2 \wedge
\theta^5 \ , \ \ \ \ \ \ \ \ \ \ 
          \diff \bar{\Theta}{}^2 \=  -2 \bar{\Theta}{}^2 \wedge \theta^6 - 2
          \bar{\Theta}{}^2 \wedge \theta^5 \ , \\[4pt]
         \diff \theta^5 &=& -\mbox{$\frac{3}{4}$} \lb
         \bar{\Theta}{}^1\wedge\Theta^1   + \bar{\Theta}{}^2 \wedge
         \Theta^2\rb \ , \ \ \ \ \ \ \ \ \ \ \
\diff \theta^6 \= \mbox{$\frac{3}{4}$} \lb \bar{\Theta}{}^1 \wedge \Theta^1
- \bar{\Theta}{}^2\wedge \Theta{}^2  \rb \ ,
\eea
where we denote $\theta^5 \coloneqq \frac{3\im}{4}\, e^5$ and
$\theta^6 \coloneqq a$. This yields the structure constants
\bea
\label{eq:f_const}
\nonumber f_{61}^1&=& -2 \ , \ \ \ f_{51}^1\=2 \ , \ \ \
f_{6\bar{1}}^{\bar{1}}\= 2 \ , \ \ \ f_{5\bar{1}}^{\bar{1}}\= -2 \ , \
\ \ f_{1\bar{1}}^6\= \mbox{$\frac{3}{4}$} \ , \ \ \
f_{1\bar{1}}^5\=-\mbox{$\frac{3}{4}$} \ , \\[4pt]
           f_{62}^2&=& 2 \ , \ \ \ f_{52}^2\=2 \ , \ \ \
           f_{6\bar{2}}^{\bar{2}}\= -2 \ , \ \ \
           f_{5\bar{2}}^{\bar{2}}\= -2 \ , \ \ \ f_{2\bar{2}}^6\=
           -\mbox{$\frac{3}{4}$} \ , \ \ \
           f_{2\bar{2}}^5\=-\mbox{$\frac{3}{4}$} \ .
\eea

\subsection{Graded connections}
For the graded connection \eqref{eq58}, the non-vanishing contributions to its field strength are given by
\bea
\nonumber \cf^{i\a,i\a}&=& \diff \ca^{i\a,i\a}+\ca^{i\a,i\a} \wedge \ca^{i\a,i\a} + \ca^{i\,\a,i+1\,\a} \wedge \ca^{i+1\,\a,i\,\a} +\ca^{i\,\a,i-1\,\a}\wedge \ca^{i-1\,\a,i\,\a}\\
\nonumber               && +\, \ca^{i\,\a,i\,\a +1}  \wedge \ca^{i \,\a +1,i\,\a} +\ca^{i\,\a,i\,\a -1} \wedge \ca^{i \,\a -1,i\,\a}\\[4pt]
\nonumber              &=& F^{i\a} + D \phi^{(3)}_{i\a} \wedge e^5 \\
&& +\, \big(\phi^{(1)}_{i+1\,\a}\,
 \phi^{(1)}_{i+1\,\a}{}^\dag-\phi^{(1)\,\dagger}_{i\a}\,
 \phi^{(1)}_{i\a} -\im \phi^{(3)}_{i\a}-\mbox{$\frac{3}{4}$}\,
 c_{i\a}\, \id_{k_{i\a}} \big)\, \Theta^1 \wedge \bar{\Theta}{}^1\\ \nonumber
                        &&+\, \big( \phi^{(2)}_{i\,\a +1}\,
                        \phi^{(2)}_{i\,\a +1}{}^\dag
                        -\phi^{(2)\,\dagger}_{i\a}\, \phi^{(2)}_{i\a}
                        -\im \phi^{(3)}_{i\a}+\mbox{$\frac{3}{4}$}\,
                        c_{i\a} \, \id_{k_{i\a}} \big)\, \Theta^2 \wedge
                        \bar{\Theta}{}^2 \ , \\[4pt]
\nonumber \cf^{i\,\a,i+1\,\a}&=& \diff \ca^{i\,\a,i+1\,\a} +\ca^{i\a,i\a}\wedge \ca^{i\,\a,i+1\,\a} +\ca^{i\,\a,i+1\,\a} \wedge \ca^{i+1\,\a,i+1\,\a}\\[4pt]
\nonumber                &=& D\phi^{(1)}_{i+1\,\a} \wedge \bar{\Theta}{}^1\\ 
                    &&+\, \big( \phi^{(1)}_{i+1\,\a}\,
                    \phi^{(3)}_{i+1\,\a}-\phi^{(3)}_{i\,\a}\phi^{(1)}_{i+1\,\a}-\mbox{$\frac{3\im}{2}$}
                    \, \phi^{(1)}_{i+1\,\a}\big)\, \bar{\Theta}{}^1 \wedge e^5 
                        \= - \lb \cf^{i+1\,\a,i\,\a}\rb^{\dagger} \ , \\[4pt]
\nonumber \cf^{i\,\a,i\,\a +1} &=& \diff \ca^{i\,\a,i\,\a +1} + \ca^{i\a,i\a} \wedge \ca^{i\,\a,i\,\a +1} +\ca^{i\,\a,i\,\a +1} \wedge \ca^{i\,\a +1,i\,\a +1}\\[4pt]
\nonumber                 &=& D \phi^{(2)}_{i\,\a
  +1} \wedge \bar{\Theta}{}^2\\
                          && +\, \big( \phi^{(2)}_{i\,\a +1}\, \phi^{(3)}_{i\,\a +1}- \phi^{(3)}_{i\a}\, \phi^{(2)}_{i\,\a +1}-\mbox{$\frac{3\im}{2}$}\, \phi^{(2)}_{i\,\a +1} \big)\, \bar{\Theta}{}^2 \wedge e^5
                           \= - \lb \cf^{i\,\a +1,i\,\a}\rb^{\dagger}
                           \ , \\[4pt]      
\nonumber  \cf^{i\,\a,i+1\,\a+1} &=& \ca^{i\,\a,i+1\,\a} \wedge \ca^{i+1\,\a,i+1\,\a +1} +\ca^{i\,\a,i\,\a +1} \wedge \ca^{i\,\a +1,i+1\,\a +1}\\[4pt]
        &=& \big( \phi^{(1)}_{i+1\,\a} \, \phi^{(2)}_{i+1\,\a +1} -
        \phi^{(2)}_{i\,\a +1} \, \phi^{(1)}_{i+1\,\a +1}\big)\,
        \bar{\Theta}{}^1 \wedge \bar{\Theta}{}^2 \= - \lb \cf^{i+1\,\a
          +1,i\,\a}\rb^{\dagger} \ , \\[4pt]
\nonumber \cf^{i\,\a,i+1\,\a-1} &=& \ca^{i\,\a,i+1\,\a} \wedge \ca^{i+1\,\a,i+1\,\a-1} + \ca^{i\,\a,i\,\a-1} \wedge \ca^{i\,\a-1,i+1\,\a-1} \\[4pt]
       &=& \big( \phi^{(2)\dagger}_{i\a}\, \phi^{(1)}_{i+1\,\a -1}
       -\phi_{i+1\,\a}^{(1)} \, \phi^{(2)}_{i+1\, \a}\big)\,
       \bar{\Theta}{}^1 \wedge \Theta^2 \= - \lb
       \cf^{i+1\,\a-1,i\,\a}\rb \ .
\eea
In order to evaluate the sum $\cf_{\hat{\m}\hat{\n}}\,
\cf^{\hat{\m}\hat{\n}}$, note that the metric on $T^{1,1}$ with respect to the forms
$(\Theta^1,\bar{\Theta}{}^1,\Theta^2,\bar{\Theta}{}^2,e^5)$ reads 
$g= \delta_{\m\n}\,e^{\m} \otimes e^{\n}= \Theta^1 \otimes \bar{\Theta}{}^1 +\Theta^2 \otimes \bar{\Theta}{}^2 +e^5 \otimes e^5$. Consequently, one obtains
\bea
\nonumber \cf_{\hat{\m}\hat{\n}}\, \cf^{\hat{\m}\hat{\n}}&=& \cf_{ab}
\,\cf^{ab}+2 \big( \cf_{a1}\, \cf^{a1} + \cf_{a\bar{1}}\, \cf^{a\bar{1}}
 +\cf_{a2}\, \cf^{a2} + \cf_{a\bar{2}}\, \cf^{a\bar{2}} +\cf_{a5}\,
 \cf^{a5} + \cf_{1\bar{1}}\, \cf^{1\bar{1}} + \cf_{12}\, \cf^{12} \\
\nonumber   && + \, \cf_{1\bar{2}}\, \cf^{1\bar{2}} + \cf_{15}\,
\cf^{15}+\cf_{\bar{1} 2}\, \cf^{\bar{1} 2} +\cf_{\bar{1} \bar{2}}\,
\cf^{\bar{1} \bar{2}} +\cf_{\bar{1} 5}\, \cf^{\bar{1} 5}
    + \cf_{2\bar{2}}\, \cf^{2\bar{2}} +\cf_{25}\, \cf^{25}+
    \cf_{\bar{2} 5}\, \cf^{\bar{2} 5}\big)\\[4pt]
\nonumber    &=& \cf_{ab} \, \cf^{ab} + 4 g^{ab} \lb \cf_{a1}\,
\cf_{b\bar{1}} +\cf_{a\bar{1}}\, \cf_{b 1}+\cf_{a2}\, \cf_{b\bar{2}}
+\cf_{a\bar{2}}\, \cf_{b 2}\rb 
           +2g^{ab}\, \cf_{a5}\, \cf_{b5}\\
\nonumber &&+\,8 \lb\cf_{1\bar{1}}\, \cf_{\bar{1} 1}+\cf_{12}\,
\cf_{\bar{1} \bar{2}}+\cf_{1\bar{2}}\, \cf_{\bar{1} 2}+\cf_{\bar{1}
  2}\, \cf_{1 \bar{2}}+\cf_{\bar{1} \bar{2}}\, \cf_{12}
             + \cf_{2\bar{2}}\, \cf_{\bar{2} 2}\rb \\
     &&+\, 4 \lb\cf_{15}\, \cf_{\bar{1} 5}+\cf_{\bar{1} 5}\,
     \cf_{15}+\cf_{25}\, \cf_{\bar{2} 5}+\cf_{\bar{2} 5}\, \cf_{25}\rb
     \ .
\eea
Inserting the expressions for the field strength into this term and
taking care of the correct index structure leads to the action functional (\ref{eq68}). 
Setting $\phi^{(3)}=-\frac{3\im}{4}\, \big( \Upsilon^{(1)}+\Upsilon^{(2)}\big)$
then yields
\small
\bea
\nonumber
 S_{\mathrm{r}} &=& \frac{16\pi^3}{27}\, \int_{M^d}\, \diff^d x \ \sqrt{\det(g_{M^d})} \
 \sum_{i=0}^{m_1} \ 
                      \sum_{\a =0}^{m_2}\, \mathrm{tr}
                      \left \{ \mbox{$\frac{1}{4}$} \, F_{ab}^{i\a}
                        \,^{\dagger} \, F^{i\a \, ab} + \big(
  D_{a}\phi^{(1)}_{i\a}\big)^{\dagger} \, \big( D^{a}\phi^{(1)}_{i\a}\big) + \big( D_{a}\phi^{(2)}_{i\a}\big)^{\dagger}
                          \, \big( D^{a}\phi^{(2)}_{i\a}\big)
                      \right.\\
\nonumber        &&\left.+\, \big( D_{a}\phi^{(1)}_{i+1\, \a}\big)\,
                          \big( D^{a}\phi^{(1)}_{i+1\, \a}\big)^{\dagger}
                          + \big(
                          D_{a}\phi^{(2)}_{i\, \a +1}\big)\, \big(
                          D^{a}\phi^{(2)}_{i\, \a +1}\big)^{\dagger}
                          \right.\\
\nonumber        &&\left.+\, 2 \big|\phi^{(1)}_{i\a}\,^{\dagger} \,
  \phi_{i\a}^{(1)} - \phi^{(1)}_{i+1 \, \a} \, \phi^{(1)}_{i+1 \, \a}{}^{\dagger}
                         +\mbox{$\frac{3}{2}$}\,
                         (m_1-2i) \, \id_{k_{i\a}}\big|^2 \right. \\
\nonumber       &&\left. +\, 2 \big| \phi^{(2)}_{i\a}\,^{\dagger} \, \phi_{i\a}^{(2)} - \phi^{(2)}_{i+1\, \a} \, \phi^{(2)}_{i+1\, \a}{}^{\dagger}
                         +\mbox{$\frac{3}{2}$}\,
                         (m_2-2\a) \, \id_{k_{i\a}}\big|^2 \right. \\
\nonumber       &&\left. +\, 2 \big|\phi^{(1)}_{i+1\, \a}\,
  \phi^{(2)}_{i+1\, \a +1}-\phi^{(2)}_{i\, \a +1} \, \phi^{(1)}_{i+1\, \a +1} \big|^2
                     + 2 \big|\, \big(\phi^{(1)}_{i\, \a -1} \, \phi^{(2)}_{i\a}-
                     \phi^{(2)}_{i-1\, \a}\, \phi^{(1)}_{i\a} \big)^\dag
                     \, \big|^2 \right.\\
\label{SYM_red2}      && \left. +\, 2 \big| \phi^{(2)}_{i\a} \,^{\dagger} \,
  \phi^{(1)}_{i+1\, \a -1} - \phi^{(1)}_{i+1\, \a} \,
  \phi^{(2)}_{i+1\, \a}{}^{\dagger}\big|^2 + 2 \big| \, \big(
  \phi^{(2)}_{i-1\, \a
    +1}{}^{\dagger} \, \phi^{(1)}_{i\a}
                    -\phi^{(1)}_{i\, \a +1}\, \phi^{(2)}_{i\, \a
                      +1}{}^{\dagger}\big)^\dag\, \big|^2 \right\} \ . 
\eea
\normalsize
This result coincides with the action functional derived in
\cite{LPS06} if one rescales the Higgs fields $\phi^{(1)}$ and
$\phi^{(2)}$ by the factor $\sqrt{{3}/{2}}$, which is necessary for
the comparison as $\Theta^l= \sqrt{{2}/{3}} \ \beta_l$ for $l=1,2$
in the limit $\vp=0$. Then one sees again that the Sasaki-Einstein condition has fixed the radii to $R_1^2=R_2^2=\frac16$. 

\subsection{Connections on the conifold}
From the Sasaki-Einstein condition $\diff e^5 = -2 \omega^3 = -2 \lb
e^{12}+e^{34}\rb$ it follows that the 2-form $\Omega$ is closed, as a
simple calculation shows
\bea
\nonumber \diff \Omega &=& \diff \left[ r^2 \lb \omega^3
  +e^{56}\rb\right]\\[4pt] \nonumber
                       &=& 2 r \, \diff r \wedge \omega^3 + r^2 \lb
                       -2\omega^3 \rb \wedge e^6 \\[4pt] &=& 2 r \,
                       \diff r \wedge \omega^3 - 2r \, \omega^3 \wedge
                       \diff r\=0 \ .
\eea
It induces the complex structure with $Je^5=-e^6$, which yields the holomorphic \mbox{1-form} $\Theta^3 \coloneqq e^5+\im e^6$. 
The structure equations for the rescaled forms on the cone $\et^{\m}=r
\, e^{\m}$ read
\bea
\nonumber \diff \et^1 &=& -\mbox{$\frac{3}{2r}$}\, \et^{25}-2 \im
\et^2 \wedge a -\mbox{$\frac{1}{r}$}\, \et^{16} \ , \ \ \ \ \ \ 
         \diff \et^{2}\= \mbox{$\frac{3}{2r}$}\, \et^{15}+2\im \et^1
         \wedge a -\mbox{$\frac{1}{r}$}\, \et^{26} \ , \\[4pt]
\nonumber \diff \et^3 &=& -\mbox{$\frac{3}{2r}$}\, \et^{45}+2 \im
\et^4 \wedge a -\mbox{$\frac{1}{r}$}\, \et^{36} \ , \ \ \ \ \ \ 
         \diff \et^{4}\= \mbox{$\frac{3}{2r}$}\, \et^{35}-2\im \et^3
         \wedge a -\mbox{$\frac{1}{r}$}\, \et^{46} \ , \\[4pt]
\diff \et^5 &=& -\mbox{$\frac{2}{r}$}\, \et^{12}
-\mbox{$\frac{2}{r}$}\, \et^{34}-\mbox{$\frac{1}{r}$}\, \et^{56} \ , \
\ \ \ \ \  \ \ \ \ \ \ \diff \et^6\=0 \ ,
\eea
which can be expressed in terms of complex forms as in
\eqref{gamma_cone}.
The curvature of the corresponding Levi-Civita connection $\Gamma$ is given by
\beq
R\= \diff \Gamma +\Gamma \wedge \Gamma \= \begin{pmatrix}
           2 \im \lb e^{12}-e^{34}\rb		&- \lb e^{13}+\e^{24}\rb +\im \lb e^{23}-e^{14}\rb	&0\\
           \lb e^{13}+e^{24}\rb +\im \lb e^{23}-e^{14}\rb                & -2\im \lb e^{12}-e^{34}\rb    &0\\
           0					&0							&0               
                                          \end{pmatrix} \ ,
\eeq
and hence it is valued in $\mathfrak{su}(2) \subset \mathfrak{su}(3)$. This curvature solves the 
instanton equation and confirms the Ricci-flatness of the metric cone
$C(T^{1,1})$ (by contracting the components of $R$ to the Ricci tensor).


\end{document}